\documentclass[]{spie}  
\usepackage[]{graphicx}
\usepackage{amsmath,amssymb,mathtools,url}
\title{Performance and sensitivity of vortex coronagraphs on segmented space telescopes} 
\usepackage{array}
\newcolumntype{P}[1]{>{\centering\arraybackslash}p{#1}}

\author{Garreth~Ruane\supit{a}, Dimitri~Mawet\supit{a,b}, Jeffrey~Jewell\supit{b}, and Stuart~Shaklan\supit{b}
\skiplinehalf
\supit{a}Department of Astronomy, California Institute of Technology, 1200 E. California Blvd.,\\Pasadena, CA 91125, USA; \\
\supit{b}Jet Propulsion Laboratory, California Institute of Technology, 4800 Oak Grove Dr.,\\Pasadena, CA 91109, USA
}

\authorinfo{Further author information: send correspondence to gruane@caltech.edu}

\graphicspath{{Figures/}}
 
\begin{document} 
  \maketitle 

\begin{abstract}
The detection of molecular species in the atmospheres of earth-like exoplanets orbiting nearby stars requires an optical system that suppresses starlight and maximizes the sensitivity to the weak planet signals at small angular separations. Achieving sufficient contrast performance on a segmented aperture space telescope is particularly challenging due to unwanted diffraction within the telescope from amplitude and phase discontinuities in the pupil. Apodized vortex coronagraphs are a promising solution that theoretically meet the performance needs for high contrast imaging with future segmented space telescopes. We investigate the sensitivity of apodized vortex coronagraphs to the expected aberrations, including segment co-phasing errors in piston and tip/tilt as well as other low-order and mid-spatial frequency aberrations. Coronagraph designs and their associated telescope requirements are identified for conceptual HabEx and LUVOIR telescope designs. 
\end{abstract}



\section{INTRODUCTION}
\label{sec:intro}  

Directly detecting atmospheric biomarkers on earth-like exoplanets orbiting sun-like stars is one of the premier science goals of future, large-aperture ($>$4~m) space telescopes. In particular, two mission concepts currently under study in preparation for the 2020 Astrophysics Decadal Survey will have exoplanet imaging and spectroscopy capabilities: the Habitable Exoplanet Imaging Mission (HabEx)\cite{Mennesson2016} and the Large UV/Optical/IR Surveyor (LUVOIR)\cite{Bolcar2016}. The HabEx architecture is a fully off-axis telescope with either a 4~m monolithic or 6.5~m segmented primary mirror. On the other hand, the LUVOIR design is a centrally-obscured 9-15~m segmented telescope. With each concept, achieving sufficient starlight suppression for imaging earth-like exoplanets with an on-board coronagraph instrument requires specially designed coronagraph masks, an ultra-stable telescope, and high-precision wavefront control. 

The coronagraph may be designed to passively reject unwanted diffraction within the telescope from amplitude discontinuities and obstructions in the pupil, including the secondary mirror, spider support structures, and gaps between mirror segments. The coronagraph can also be made robust to the most problematic low-order aberrations. Deformable mirrors are used to correct for static imperfections in optical surfaces. However, even with a state-of-the-art coronagraph and wavefront control system, the contrast performance of coronagraphs is limited by dynamic mid-spatial frequency aberrations. 

Here, we present vortex coronagraph designs for HabEx and LUVOIR that are robust to low-order aberrations. We set wavefront stability requirements on the telescope, including the phasing of the primary mirror segments, and discuss how some telescope requirements may be relaxed by trading robustness to aberrations for planet throughput. 

\section{Coronagraph performance metrics}
\label{sec:metrics}
The efficiency of coronagraph instruments will have a strong influence on the scientific yield of future missions. The best coronagraph designs for HabEx and LUVOIR provide the largest number of detected and characterized earth-like exoplanets within the mission lifetime. As a proxy, we consider an optimal coronagraph one that minimizes the exposure time needed to detect notional earth-like exoplanets in the habitable zone around nearby sun-like stars. 

The signal from the planet and star measured in a region-of-interest at the position of the planet is defined as $S_p = \eta_p \Phi_p \Delta t \Delta\lambda A q T$ and $S_s = \eta_s \Phi_s \Delta t \Delta\lambda A q T$, where $\eta_p$ and $\eta_s$ are the fraction of planet and star light detected, $\Phi_p$ and $\Phi_s$ are the flux owing to the planet and star (photons per unit area per unit time per unit wavelength at the primary mirror), $\Delta t$ is the integration time, $\Delta\lambda$ is the spectral bandwidth, $A$ is the collecting area of the telescope, $q$ is the detective quantum efficiency, and $T$ is the transmission of the instrument describing losses that affect the star and planet equally. For the purpose of this work, the coronagraph performance is described by $\eta_p$ and $\eta_s$, which both depend on bandwidth, angular separation from the star, and angular extent of the star. When limited by photon noise from diffracted starlight, the signal-to-noise ratio (SNR) is given by
\begin{equation}
\mathrm{SNR} = \frac{S_p}{\sqrt{S_s}}=\frac{\eta_p }{\sqrt{\eta_s}}\frac{\Phi_p}{\sqrt{ \Phi_s}}\sqrt{\Delta t \Delta\lambda A q T}.
\label{eq:snr1}
\end{equation}
It is common to define a raw contrast ratio as $C = \eta_s/\eta_p$ and the astrophysical contrast (or flux) ratio as 
$\epsilon = \Phi_p/\Phi_s$. 
Solving for the exposure time to achieve a given SNR, $\Gamma$, we arrive at two equivalent expressions:
\begin{align}
\Delta t &=  \frac{\eta_s}{\eta_p^2}\frac{\Phi_s  }{\Phi_p^2}\frac{\Gamma^2 }{ \Delta\lambda A q T} \\
&=  \frac{C}{\eta_p} \frac{1}{\epsilon^2 \Phi_s}\frac{\Gamma^2}{\Delta\lambda A q T}.
\end{align}
An optimal coronagraph minimizes $\Delta t$, which is proportional to $\eta_s/\eta_p^2$ or $C/\eta_p$ in the photon-noise-limited regime (note that $C$ itself is a function of $\eta_p$). 

In practice, many other significant noise sources will be present and the SNR will take the form
\begin{equation}
    \mathrm{SNR}=\frac{S_p}{\sqrt{S_s+\sigma^2}},
\end{equation}
where $\sigma^2$ represents the sum of the variances for all additional noise sources. The dominant terms fall into two categories, where the variance is approximately proportional to $\Delta t$ (e.g. detector noise and static background) or $\Delta t^2$ (e.g. residual spatial speckle noise). We therefore expand the noise variance as follows: $\sigma^2=n_1 \Delta t \Phi_s \Delta\lambda q A T + (n_2 \Delta t \Phi_s \Delta\lambda q A T)^2 $, where $n_1$ and $n_2$ describe the effective strength of the additional noise sources. With this, the expression for exposure time is
\begin{equation}
\Delta t =  \left[\frac{\eta_s + n_1}{\eta_p^2 - (n_2 \Gamma/\epsilon)^2}\right]\frac{1}{\epsilon^2\Phi_s}\frac{\Gamma^2 }{ \Delta\lambda A q T}
\end{equation}
and the optimal coronagraph satisfies 
\begin{equation}
\min_{\eta_p,\eta_s} \left[\frac{\eta_s + n_1}{\eta_p^2 - (n_2 \Gamma/\epsilon)^2}\right],\text{ subject to: } \eta_p>n_2\Gamma/\epsilon.
\end{equation}
Again, in the regime where the performance is limited by photon noise from diffracted starlight (i.e. $n_1,n_2\approx0$), $\Delta t \propto \eta_s/\eta_p^2$. When limited by another noise source whose variance scales with $\Delta t$, i.e. $n_1 \gg \eta_s$, the performance of a coronagraph design only improves by increasing throughput, $\eta_p$. The majority of contributions to $n_2$ will be due to dynamic mid-spatial frequency errors, which are not passively suppressed by the coronagraph. Therefore, the throughput must also be high enough to maintain $\eta_p > n_2 \Gamma/\epsilon$, where $\epsilon\approx10^{-11}\text{-}10^{-10}$. In fact, raw contrast values reported from laboratory testing and goals laid out in space mission milestones often refer to ensuring $n_2/\eta_p < \epsilon/\Gamma$. While wavefront control and differential imaging are used to minimize $n_2$, the coronagraph must also be designed to maximize $\eta_p$ in order to ensure detection and minimize the required exposure time. In the following, we report the influence of the coronagraph masks and optical aberrations in the telescope in terms of $\eta_s$ and $\eta_p$.


\section{OFF-AXIS, UNOBSCURED TELESCOPE} \label{sec:monolith} 

\begin{figure}[t]
    \centering
    \includegraphics[height=5.25cm,trim={2.5cm -1cm 0 0},clip]{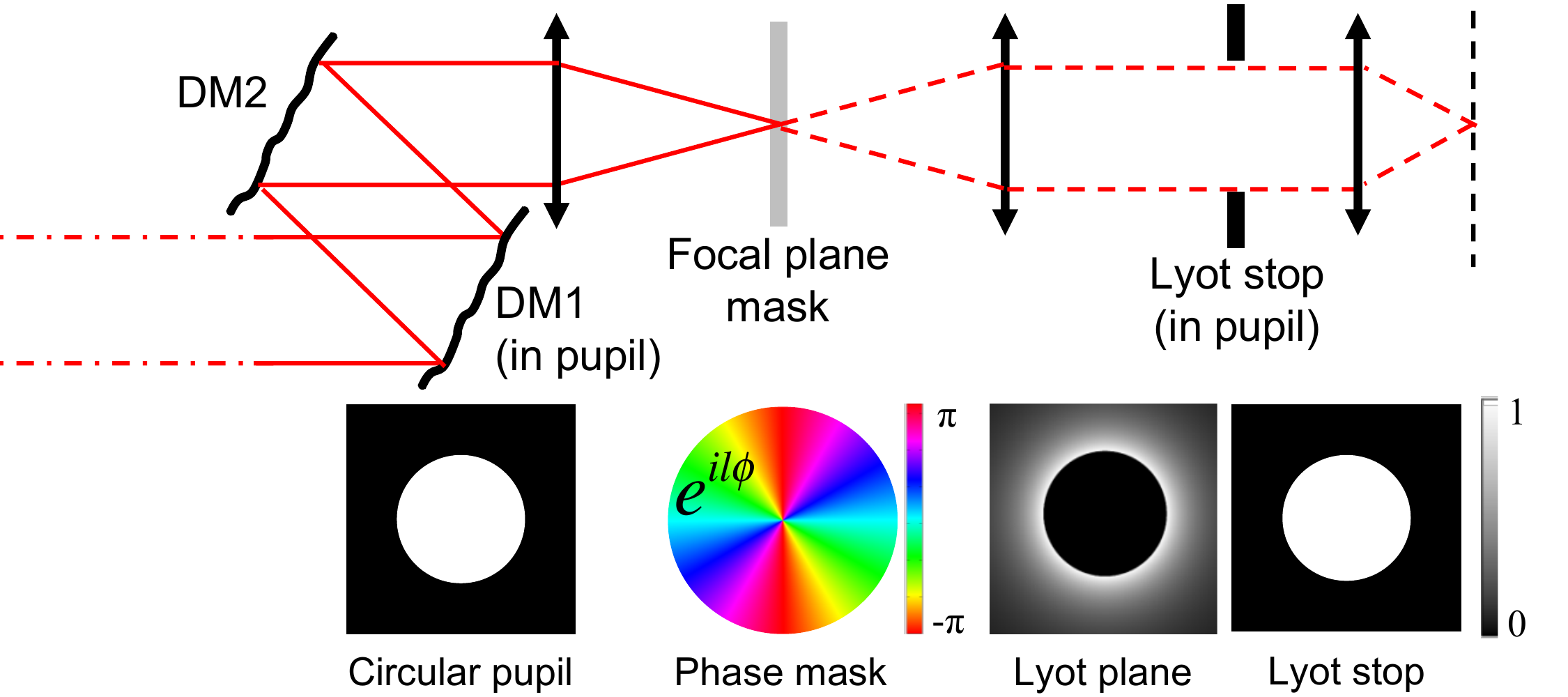}
    \includegraphics[height=5.5cm]{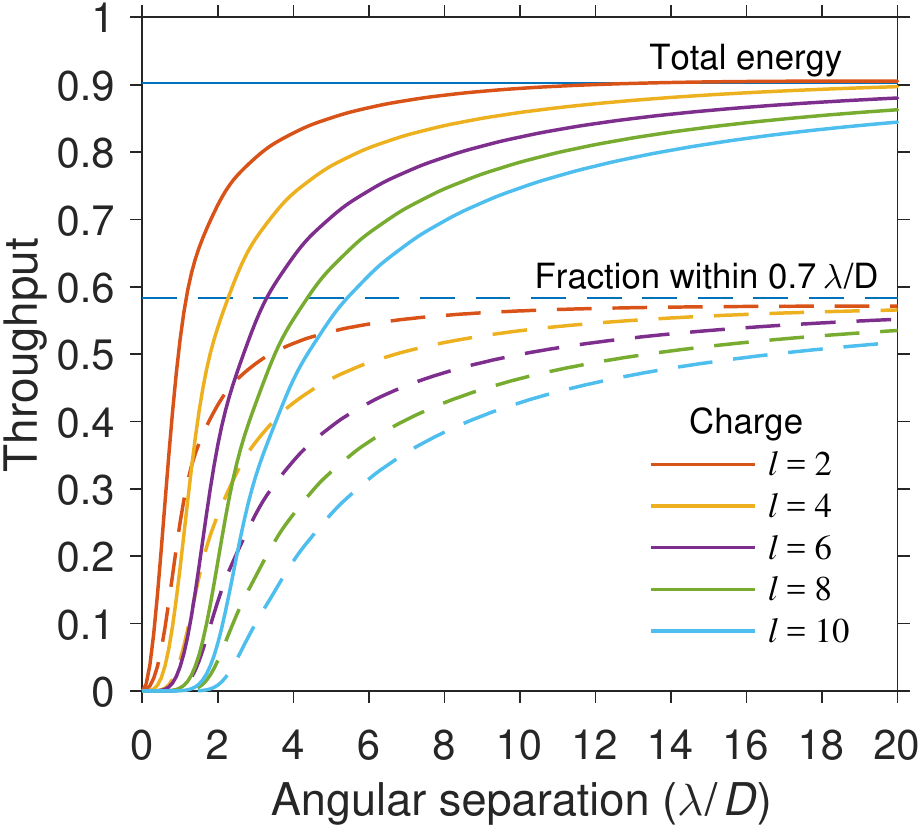}
    \caption{\textit{left:}~Schematic of a vortex coronagraph with deformable mirrors DM1 and DM2, focal plane phase mask with complex transmittance $\exp(il\phi)$, and circular Lyot stop. Starlight suppression is achieved by diffracting the stellar field outside of the Lyot stop. \textit{right:}~Throughput performance of a vortex coronagraph for $b/a=0.95$. }
    \label{fig:VCschematic}
\end{figure}

The first telescope architecture we consider is a 4~m off-axis HabEx concept with a monolithic primary mirror. The unobstructed pupil is conducive to highly efficient coronagraph designs, such as the vortex coronagraph \cite{Mawet2005,Foo2005,Mawet2010a}. A vortex coronagraph is an optical system that suppresses starlight and provides high sensitivity to weak planet signals at small angular separations, as demonstrated in the laboratory \cite{Mawet2009} and on-sky with ground based telescopes \cite{Mawet2010b,Serabyn2010,Mawet2017,Serabyn2017,Ruane2017}. Figure \ref{fig:VCschematic}, \textit{left} shows a schematic of a vortex coronagraph with a dual deformable mirrors for wavefront control, a focal plane mask, and Lyot stop. The vortex focal plane mask is a transparent optic which imparts a spiral phase shift of the form $\exp(il\phi)$ on the incident field, where $l$ is a even non-zero integer known as the ``charge" and $\phi$ is the azimuth angle in the focal plane. Light from an on-axis point source (i.e. the star) that passes through the circular entrance pupil of radius $a$ is completely diffracted outside of the downstream Lyot stop of radius $b$, assuming $b<a$ and one-to-one magnification within the coronagraph. In addition to ideal starlight suppression, the vortex coronagraph provides high throughput for point-like sources at small angular separations from the star (see Fig. \ref{fig:VCschematic}, \textit{right}).

\subsection{Ideal throughput of vortex coronagraphs}

Unlike coronagraphs with occulting masks, the throughput of a vortex coronagraph is a smooth function of the planet's angular separation, which increases more slowly for higher values of $l$, and ideally approaches unity at large angular separations. However, an undersized Lyot stop radius is used to improve robustness to pupil alignment errors. Also, a smaller Lyot stop and higher $l$ value both reduce sensitivity to low order aberrations.

The specific definition of throughput varies in the literature. Here, we present two common definitions: (1)~the fraction of planet energy from a planet that reaches the image plane and (2)~the fraction of the planet energy that falls within a circular region-of-interest with radius $\hat{r}\lambda/D$ centered at the planet position, where $\lambda$ is the wavelength and $D$ is the diameter of the primary mirror. The maximum throughput (at large angular separations) in each case is
\begin{equation}
    \eta_{p,\text{max}}=
    \begin{cases}
        (b/a)^2, & \text{(1) total energy}\\
        (b/a)^2\left[1-J_0\left(\pi \hat{r}  \frac{b}{a}\right)^2-J_1\left(\pi \hat{r} \frac{b}{a}\right)^2\right], & \text{(2) fraction of total energy within a } \hat{r} \lambda/D \text{ radius}\\
    \end{cases},
\end{equation}
where $J_0(~)$ and $J_1(~)$ are Bessel functions of the first kind. For example, if $b/a=0.95$ and $\hat{r}=0.7$, the theoretical maxima for case (1) and (2) are 90\% and 58\%, respectively. The latter value may also be normalized to the same quantity without the coronagraph masks. For example, in the case described above, 86\% of planet energy remains within 0.7$\lambda/D$ of the planet's position in the image, a value referred to as the relative throughput. Definitions (1) and (2) are plotted for various values of $l$ in Fig. \ref{fig:VCschematic}, \textit{right}.

\subsection{Passive insensitivity to low-order aberrations}

Detecting earth-like exoplanets in practice will require a coronagraph whose performance is insensitive to wavefront errors owing to mechanical motions in the telescope and differential polarization aberrations, which both manifest as low-order aberrations. We describe the phase at the entrance pupil of the coronagraph as a linear combination of Zernike polynomials $Z_n^m (r/a,\theta)$ defined over a circular pupil of radius $a$. An isolated phase aberration is written
\begin{equation}
P(r,\theta) = \exp \left[i c_{nm} Z_n^m(r/a,\theta)\right],\;\;\;\;\;r\le a,
\end{equation}
where $i=\sqrt{-1}$ and $c_{n,m}$ is a coefficient. Assuming small wavefront errors (i.e. $c_{nm}\ll$ 1 rad), the field in the pupil may be approximated to first order via its Taylor series expansion:
\begin{equation}
P(r,\theta) \approx 1 + i c_{nm} Z_n^m(r/a,\theta),\;\;\;\;\;r\le a.
\label{eq:Ztaylor}
\end{equation}
The propagation of each mode through the coronagraph is described analytically in Appendix B. In cases where all of the light is located outside of the geometric pupil, the source is extinguished by the Lyot stop. The constant term in Eq. \ref{eq:Ztaylor} is completely suppressed for all nonzero even values of $l$. The first order $Z_n^m(r/a,\theta)$ term is also blocked by the Lyot stop if $|l|>n+|m|$. Through this mechanism, increasing $l$ leads to relaxed wavefront error requirements\cite{Mawet2010a,Ruane2015_SPIE}.

\subsection{Wavefront error requirements}

The coronagraph and telescope designs must be jointly optimized to passively suppress starlight and provide the wavefront stability needed to maintain suppression throughout an observation. However, the wavefront requirements for a given coronagraph design depends on the aberration mode and/or spatial frequency content of the error. Telescope requirements for earth-like exoplanet imaging with vortex coronagraphs are presented in this section in terms of low-order and mid-to-high spatial frequency aberrations.

\begin{figure}[t!]
    \centering
    \includegraphics[scale=0.9]{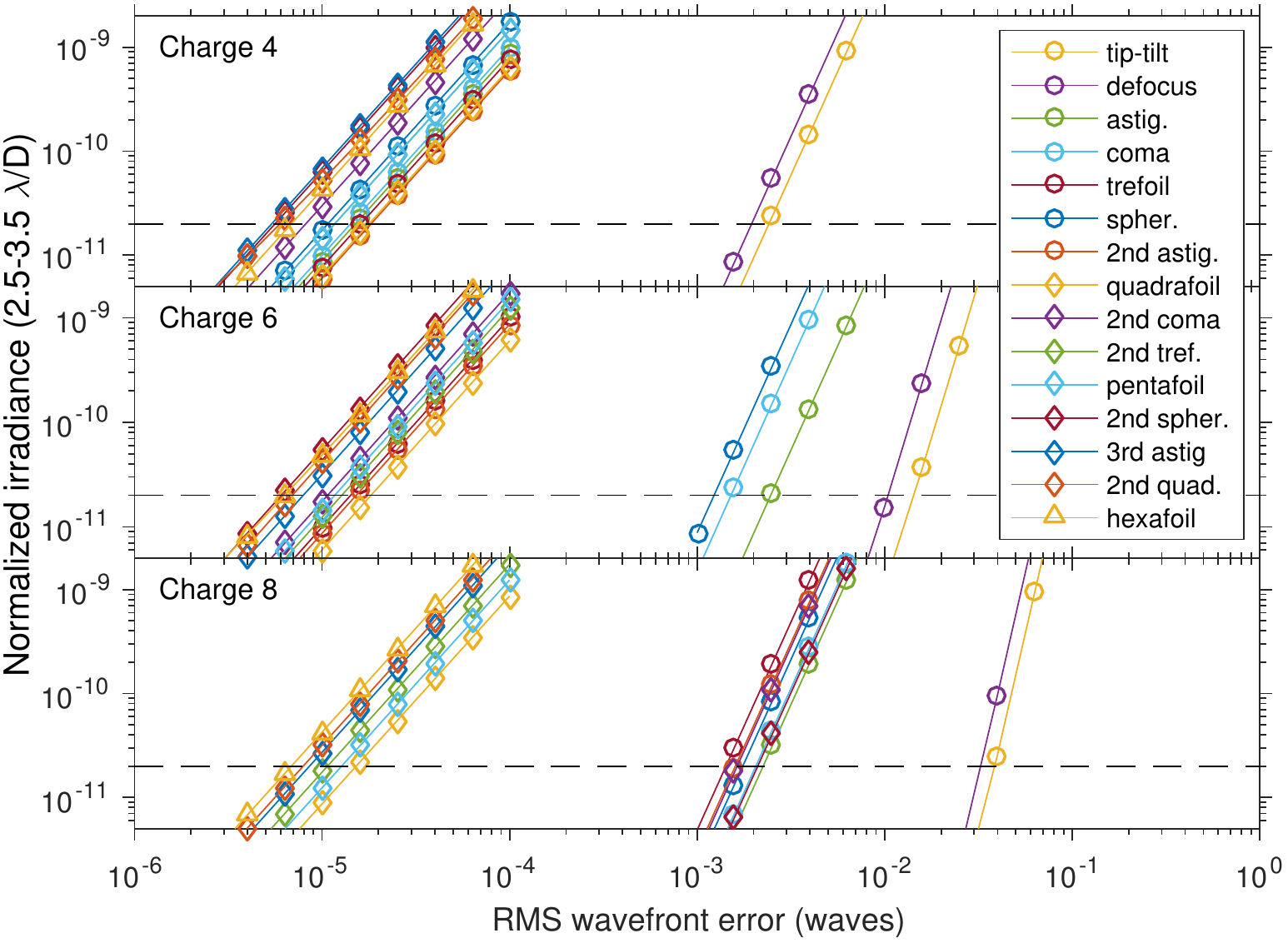}
    \caption{Sensitivity of the vortex coronagraph to low-order aberrations. Stellar irradiance averaged over angular separations 2.5-3.5~$\lambda/D$, normalized to the peak irradiance without the coronagraph masks, as a function of root-mean-square (RMS) wavefront error in each Zernike aberration.}
    \label{fig:IrrVsWFE}
\end{figure}


\subsubsection{Low order requirements: Zernike aberrations}

Figure \ref{fig:IrrVsWFE} shows the leaked starlight through the coronagraph (stellar irradiance, averaged over angular separations 2.5-3.5~$\lambda/D$, and normalized to the peak value without the coronagraph masks) as a function of root-mean-square (RMS) wavefront error. Modes with $n+|m|\ge l$ follow a quadratic power law and generate irradiance at the $\sim10^{-11}$ level for wavefront errors of $\sim10^{-5}$ waves rms. However, modes with $n+|m|<l$ are blocked at least to first order at the Lyot stop, as described in the previous section. In these cases, the equivalent irradiance level ($\sim10^{-11}$) corresponds to $\sim100\times$ the wavefront error.

We place requirements on the telescope by setting a maximum allowable irradiance threshold on the leaked starlight at $3\pm0.5~\lambda/D$. Here, the threshold is chosen to be 2$\times10^{-11}$ per Zernike mode (dashed line in Fig. \ref{fig:IrrVsWFE}). The corresponding wavefront error at $\lambda=450$~nm, likely the shortest and most challenging wavelength, are reported in Table \ref{tab:loworder}. Modes that are suppressed by the coronagraph have wavefront requirements $\gtrsim100$~pm~rms, while those that transmit tend to have require $<$10~pm~rms. The minimum charge of the vortex coronagraph may be chosen to preserve robustness to particularly problematic low-order aberrations as well as to relax requirements and reduce the cost of the overall mission. However, increasing the minimum charge has a significant impact on the scientific yield of the mission, especially since insufficient throughput at small angular separations (i.e. beyond the so-called ``inner working angle") will likely limit the number of detected and characterized earth-like planets within the mission lifetime \cite{Stark2015}.

The requirements given in Table \ref{tab:loworder} may be scaled to any wavelength by simply multiplying the reported rms wavefront error by a factor of $\lambda/(450~\text{nm})$. While a higher charge (e.g. charge 6 or 8) may be used for the shortest wavelengths to improve robustness, using a lower charge (e.g. charge 4) at longer wavelengths would allow exoplanets detected near the inner working angle of the visible coronagraph to be characterized in the infrared, where the wavefront error requirements are naturally less strict. In that case, the infrared coronagraph would drive requirements in some of the lowest order modes, which would be relaxed by a factor of $\gtrsim$2 with respect to higher-order requirements driven by the visible coronagraph.  

\begin{table}[t!]
\caption{Low-order wavefront error requirements for low-order Zernike modes at $\lambda=450~\text{nm}$.}
\label{tab:loworder}
\begin{center}       
\begin{tabular}{|c|P{0.9cm}|P{0.9cm}|P{0.9cm}|P{1.5cm}|P{1.5cm}|P{1.5cm}|P{1.5cm}|}
\hline
\rule[-1ex]{0pt}{3.5ex} Aberration & \multicolumn{3}{c}{Indices} & \multicolumn{4}{|c|}{Allowable RMS wavefront error per mode (nm) }\\
\hline
\rule[-1ex]{0pt}{3.5ex}  & Noll & $n$ & $m$ & $l=4$ & $l=6$ & $l=8$ & $l=10$\\
\hline\hline
\rule[-1ex]{0pt}{3.5ex} Tip-tilt & 2,3 & 1 & $\pm$1 & 1.1 & 5.9 & 14 & 26 \\
\hline
\rule[-1ex]{0pt}{3.5ex} Defocus & 4 & 2 & 0 & 0.81 & 4.6 & 12 & 26 \\
\hline
\rule[-1ex]{0pt}{3.5ex} Astigmatism  & 5,6 & 2 & $\pm$2 & 0.0067 & 1.1 & 0.9 & 4.6 \\
\hline
\rule[-1ex]{0pt}{3.5ex} Coma & 7,8 & 3 & $\pm$1 & 0.0062 & 0.66 & 0.82 & 5 \\
\hline
\rule[-1ex]{0pt}{3.5ex} Trefoil & 9,10 & 3 & $\pm$3 & 0.0072 & 0.0063 & 0.57 & 0.67 \\
\hline
\rule[-1ex]{0pt}{3.5ex} Spherical & 11 & 4 & 0 & 0.0048 & 0.51 & 0.73 & 6.3 \\
\hline
\rule[-1ex]{0pt}{3.5ex} $2^\text{nd}$ Astig. & 12,13 & 4 & $\pm$2 & 0.008 & 0.0068 & 0.67 & 0.73 \\
\hline
\rule[-1ex]{0pt}{3.5ex} Quadrafoil & 14,15 & 4 & $\pm$4 & 0.0078 & 0.008 & 0.0061 & 0.54 \\
\hline
\rule[-1ex]{0pt}{3.5ex} $2^\text{nd}$ Coma & 16,17 & 5 & $\pm$1 & 0.0036 & 0.0048 & 0.69 & 0.85 \\
\hline
\rule[-1ex]{0pt}{3.5ex} $2^\text{nd}$ Trefoil & 18,19 & 5 & $\pm$3 & 0.0051 & 0.0056 & 0.0043 & 0.72 \\
\hline
\rule[-1ex]{0pt}{3.5ex} Pentafoil & 20,21 & 5 & $\pm$5 & 0.0051 & 0.0051 & 0.0051 & 0.0048 \\
\hline
\rule[-1ex]{0pt}{3.5ex} $2^\text{nd}$ Spherical & 22 & 6 & 0 & 0.0025 & 0.0027 & 0.84 & 1.1 \\
\hline
\rule[-1ex]{0pt}{3.5ex} $3^\text{rd}$ Astig. & 23,24 & 6 & $\pm$2 & 0.0024 & 0.0035 & 0.0034 & 0.82 \\
\hline
\rule[-1ex]{0pt}{3.5ex} $2^\text{nd}$ Quadrafoil & 25,26 & 6 & $\pm$4 & 0.0027 & 0.0031 & 0.0032 & 0.0038 \\
\hline
\rule[-1ex]{0pt}{3.5ex} Hexafoil & 27,28 & 6 & $\pm$6 & 0.003 & 0.0028 & 0.0028 & 0.0035 \\
\hline
\end{tabular}
\end{center}
\end{table} 

\subsubsection{Mid-to-high spatial frequency requirements}

Whereas the coronagraph design provides degrees of freedom for control robustness to low-order aberrations, high throughput coronagraphs are naturally sensitive to mid- and high-spatial frequency aberrations. In fact, any coronagraph which passively suppresses mid-spatial frequency aberrations must also have low throughput for off-axis planets. This is an outcome of the well known relationship between raw contrast and the RMS wavefront error in Fourier modes\cite{Malbet1995}. The pupil field associated with a single spatial frequency is given by
\begin{equation}
P(r,\theta) = \exp \left[i 2\sqrt{2}\pi \omega \sin\left(\frac{2 \pi x}{a} \xi  \right)\right] \approx 1 + i 2\sqrt{2}\pi \omega \sin\left(\frac{2 \pi x}{a} \xi\right),\;\;\;\;\;r\le a,
\end{equation}
where $r^2=x^2+y^2$, $\xi$ is the spatial frequency in cycles per pupil diameter, and $\omega$ is the RMS phase error in waves where we have assumed $\omega \ll 1$. The corresponding field just before the focal plane mask is
\begin{equation}
F(\rho,\phi) = f_{00}(\rho,\phi)+\sqrt{2} \pi \omega \left[ f_{00}\left(\sqrt{(x^\prime-\xi\lambda F^\#)^2+y^2},\phi\right) - f_{00}\left(\sqrt{(x^\prime+\xi\lambda F^\#)^2+y^2},\phi\right) \right],
\end{equation}
where $\rho^2=x^{\prime2}+y^{\prime2}$ and $F^\# = f/(2a)$. The coronagraph completely rejects the $f_{00}(\rho,\phi)$ term. Thus, at position $(x^\prime,y^\prime)=(\xi \lambda F^\#,0)$ after the coronagraph
\begin{equation}
F(\rho,\phi) = \sqrt{2\eta_p} \pi \omega \left[ f_{00}\left(0,\phi\right) - f_{00}\left(2\xi\lambda F^\#,\phi\right)  \right].
\end{equation}
Solving for $\eta_s$, we find
\begin{equation}
\eta_s = \eta_p 2 (\pi \omega)^2 \frac{\left| f_{00}\left(0,\phi\right) - f_{00}\left(2\xi\lambda F^\#,\phi\right)  \right|}{\left| f_{00}\left(0,\phi\right) \right|}.
\end{equation}
Therefore, for $\xi \gtrsim 1$, the raw contrast at $(x^\prime,y^\prime)=(\xi \lambda F^\#,0)$ is
\begin{equation}
C = \eta_s/\eta_p \approx 2 (\pi \omega)^2.
\end{equation}
For example, a 1 pm rms mid-spatial frequency wavefront error described by the vector $\vec{\xi}=\xi_x \hat{x} + \xi_y \hat{y}$ generates a change in raw contrast of $\sim10^{-10}$ at $\lambda=450~\text{nm}$ in the corresponding image plane location $(x^\prime,y^\prime)=(\xi_x \lambda F^\# ,\xi_y \lambda F^\#)$. This implies a stability requirement of $\sim$1~pm~rms per Fourier mode for mid-spatial frequency wavefront errors.

Rejecting starlight with mid-spatial frequency phase errors and proportionally reducing the coronagraph throughput at the position of interest degrades performance in the photon-noise-limited regime where $\text{SNR}\propto\eta_p/\sqrt{\eta_s}$. In the spatial-speckle-noise-limited regime, suppressing contributions to $n_2$ only improves performance if $n_2/\eta_p$ can be reduced. Thus, an optimal coronagraph does not provide passive robustness to spatial frequencies where $(x^\prime,y^\prime)=(\xi_x \lambda F^\# ,\xi_y \lambda F^\#)$ is in the region of interest (i.e. dark hole). 

\subsection{Sensitivity to partially resolved, extended sources}

\begin{figure}[t]
    \centering
    \includegraphics[height=0.31\linewidth]{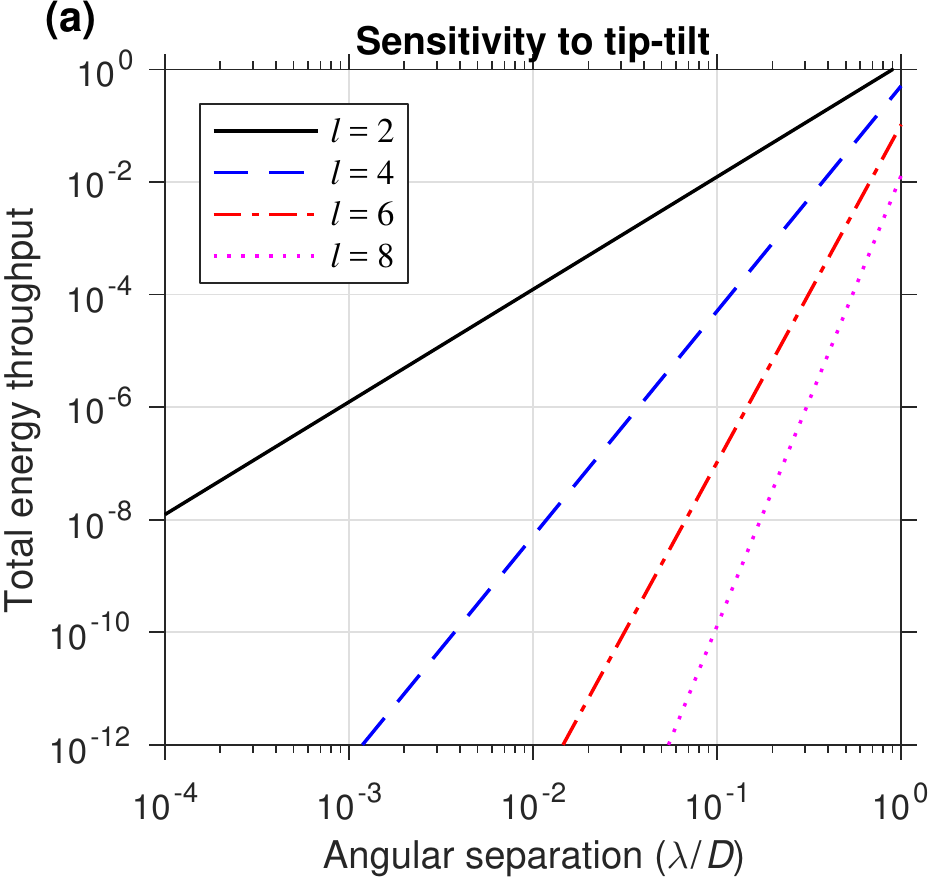}
    \includegraphics[height=0.31\linewidth]{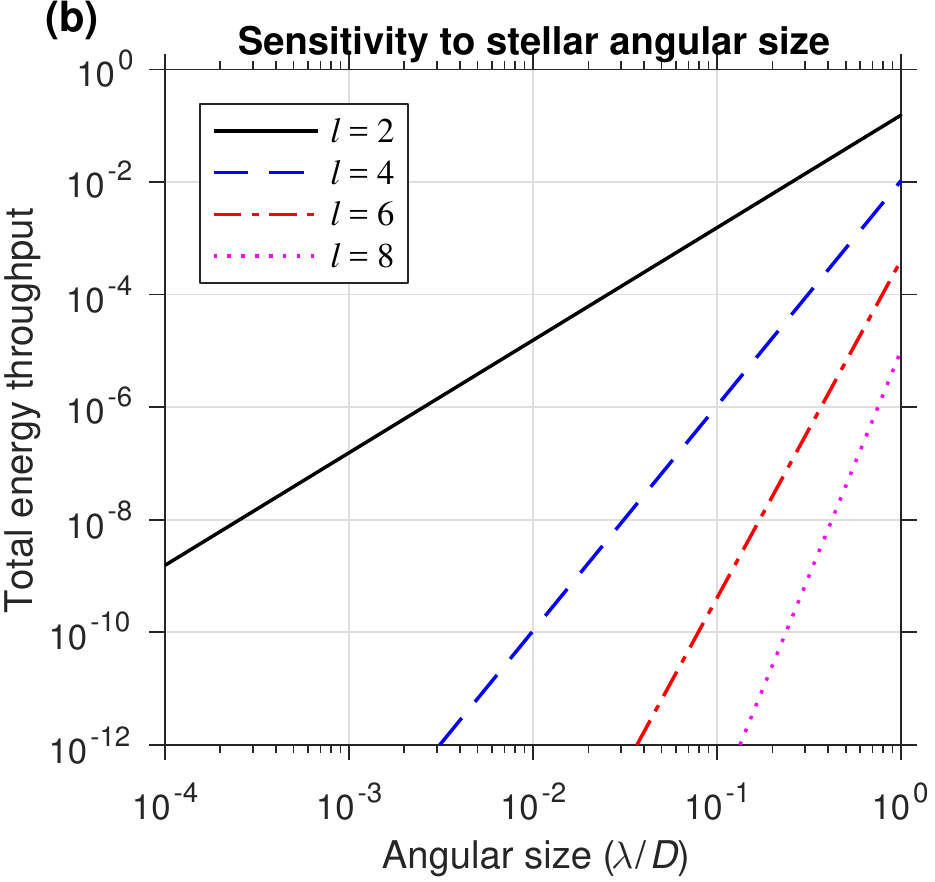}
    \caption{The fraction of stellar energy that leaks through the vortex coronagraph as a function of (a) the angular separation of a point source and (b) the angular diameter of the star.}
    \label{fig:leakfromextendedsources}
\end{figure}

The fraction of energy from a point source that leaks through the coronagraph as a function of angular separation, $\alpha$, may be approximated for small offsets (i.e. $\alpha \ll \lambda/D$) through modal decomposition of the off-axis point source in azimuth \cite{Ruane2016dissertation}. The transmitted energy is well described by a simple power law: $T_{\alpha}\propto\alpha^l$. Integrating over an extended, spatially incoherent, stellar source of angular extent, $\Theta$, leads to a similar expression: $T_{\Theta}\propto\Theta^l$. Figure \ref{fig:leakfromextendedsources} shows the leaked energy due to a point source as a function of angular separation (see Fig. \ref{fig:leakfromextendedsources}a) and as a function of stellar angular size (see Fig. \ref{fig:leakfromextendedsources}b). The stellar irradiance that appears in the image plane owing to the finite size of the star is not as critical as the contributions of dynamic mid-spatial frequency wavefront error because it is mostly incoherent with the speckles in the image plane. Therefore, we can approximate the influence on performance as only contributing to $\eta_s$, and not $n_2$. The optimal vortex charge for a given observation is simply the one that minimizes $\Delta t \propto \eta_s/\eta_p^2$ in the region of interest. 

\section{OFF-AXIS, UNOBSCURED, SEGMENTED TELESCOPE} \label{sec:offaxis_seg} 

\begin{figure}[t]
    \centering
    \includegraphics[width=\linewidth]{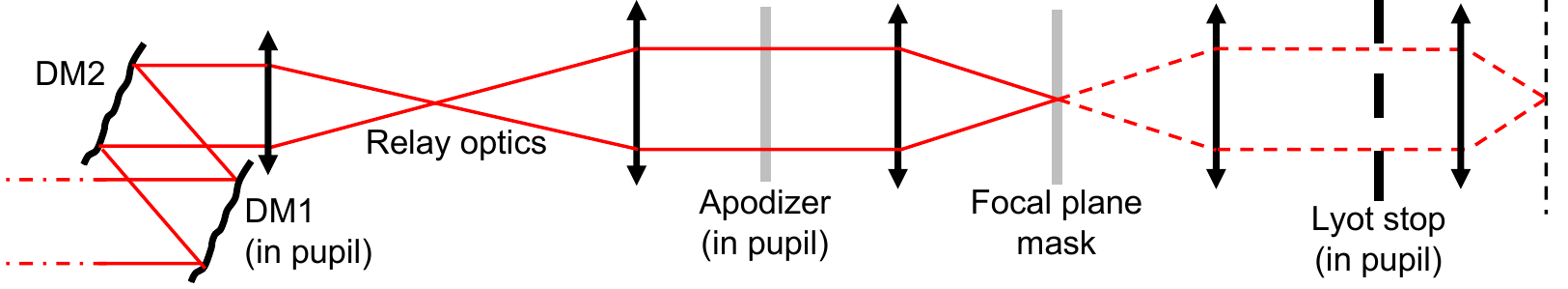}
    \caption{Schematic of an apodized vortex coronagraph. An additional relay with a gray-scale apodizer is inserted to prevent unwanted diffraction from the non-circular outer edge of the primary and gaps between mirror segments. }
    \label{fig:adodized_diagram}
\end{figure}

Another potential telescope architecture for the HabEx mission concept is a 6.5~m off-axis segmented telescope. This arrangement introduces a few additional complications with respect to the monolithic version. First, a primary mirror with a non-circular outer edge generates diffraction patterns that are difficult to null. To remedy this, we insert a circular sub-aperture in a pupil plane just before the focal plane mask, which provides improved starlight suppression at the cost of throughput (see Fig. \ref{fig:adodized_diagram}). Second, the gaps between mirror segments must be apodized to prevent unwanted diffraction in the image plane from amplitude discontinuities. In this section, we present a promising vortex coronagraph design for the 6.5~m HabEx concept and address the associated telescope requirements. 

\subsection{Apodized vortex coronagraph design}

\begin{figure}[t]
    \centering
    \includegraphics[height=0.31\linewidth]{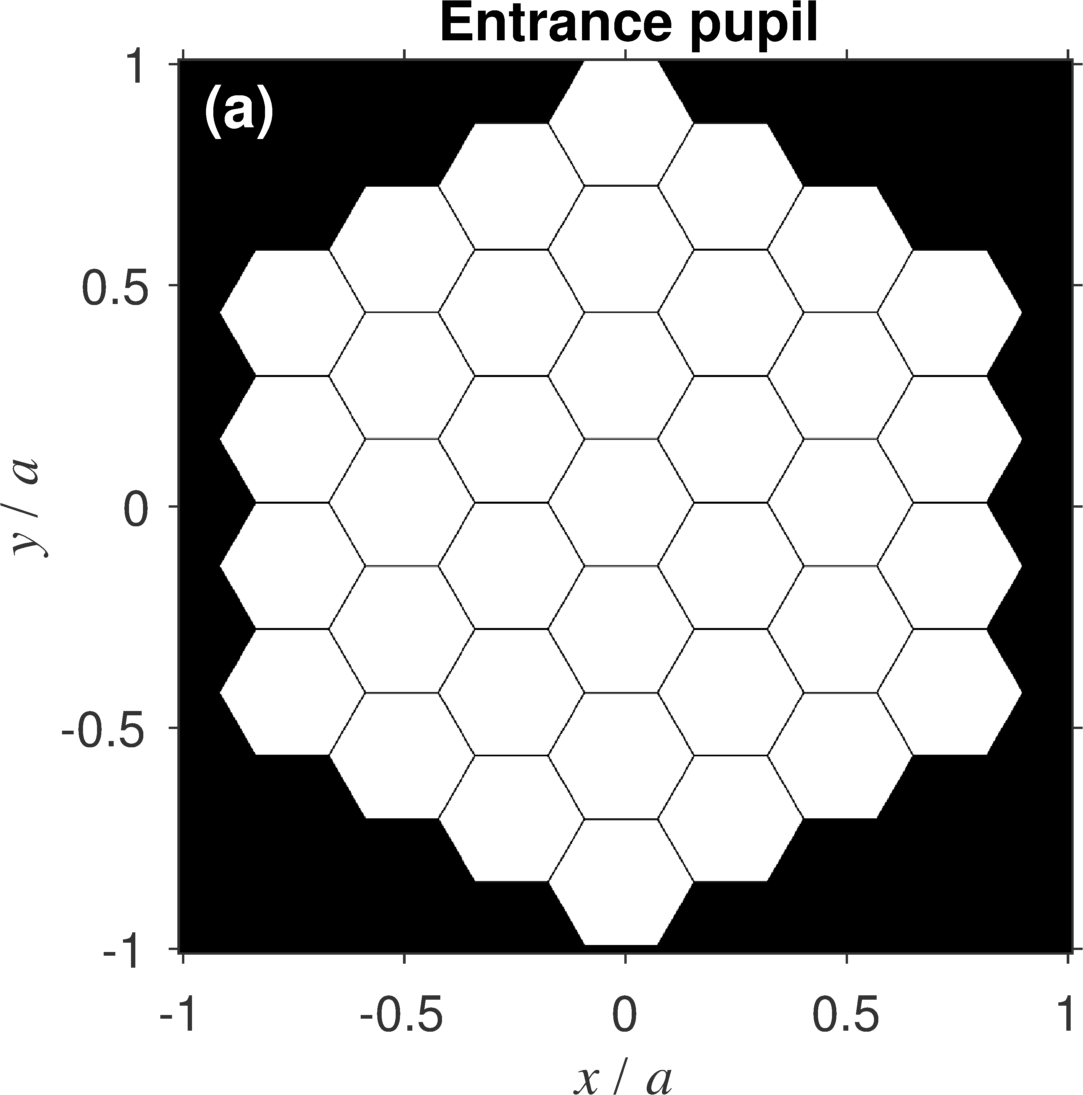}
    \includegraphics[height=0.31\linewidth,trim={1.4cm 0 0 0},clip]{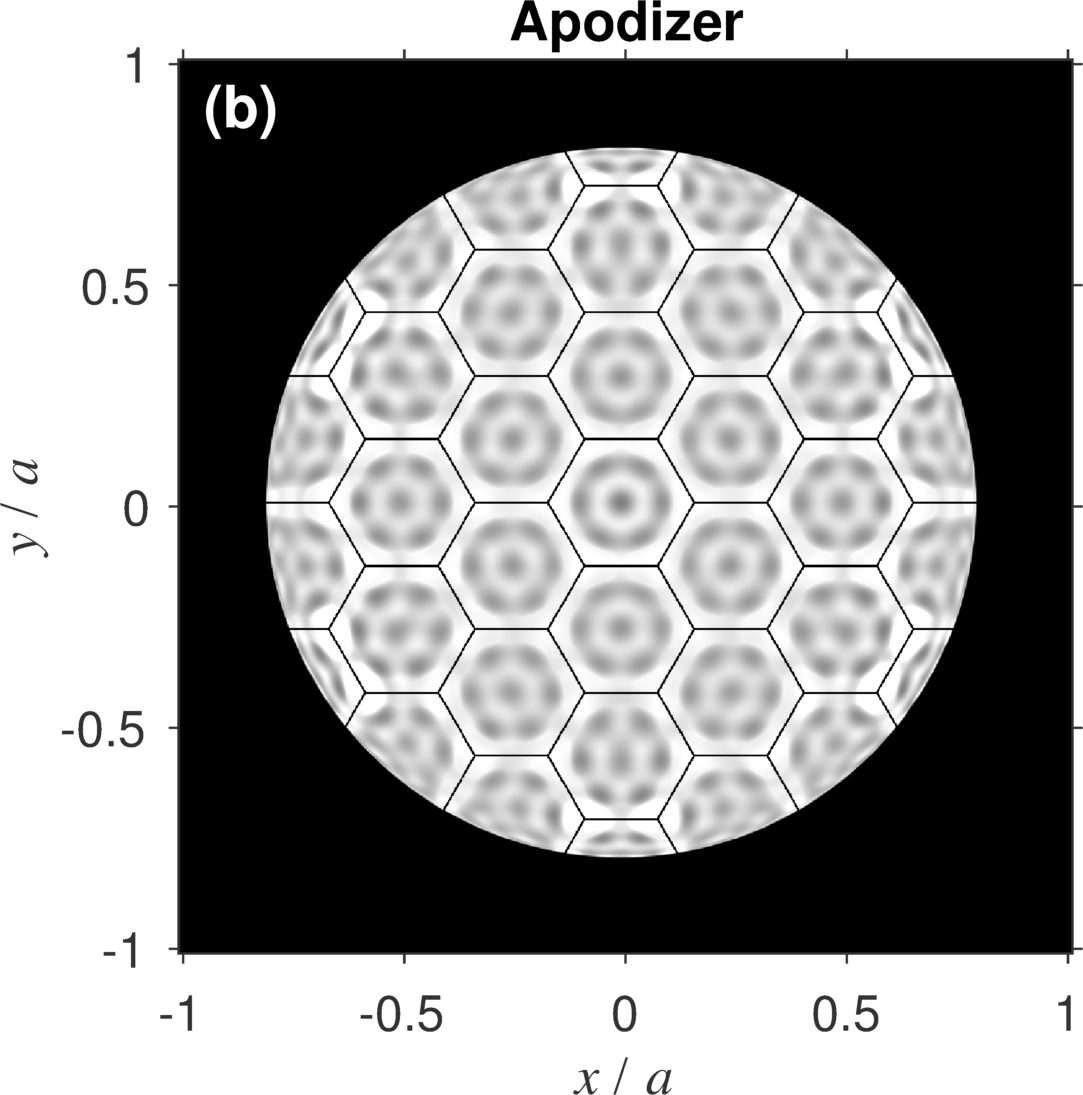}
    \includegraphics[height=0.31\linewidth,trim={1.4cm 0 0 0},clip]{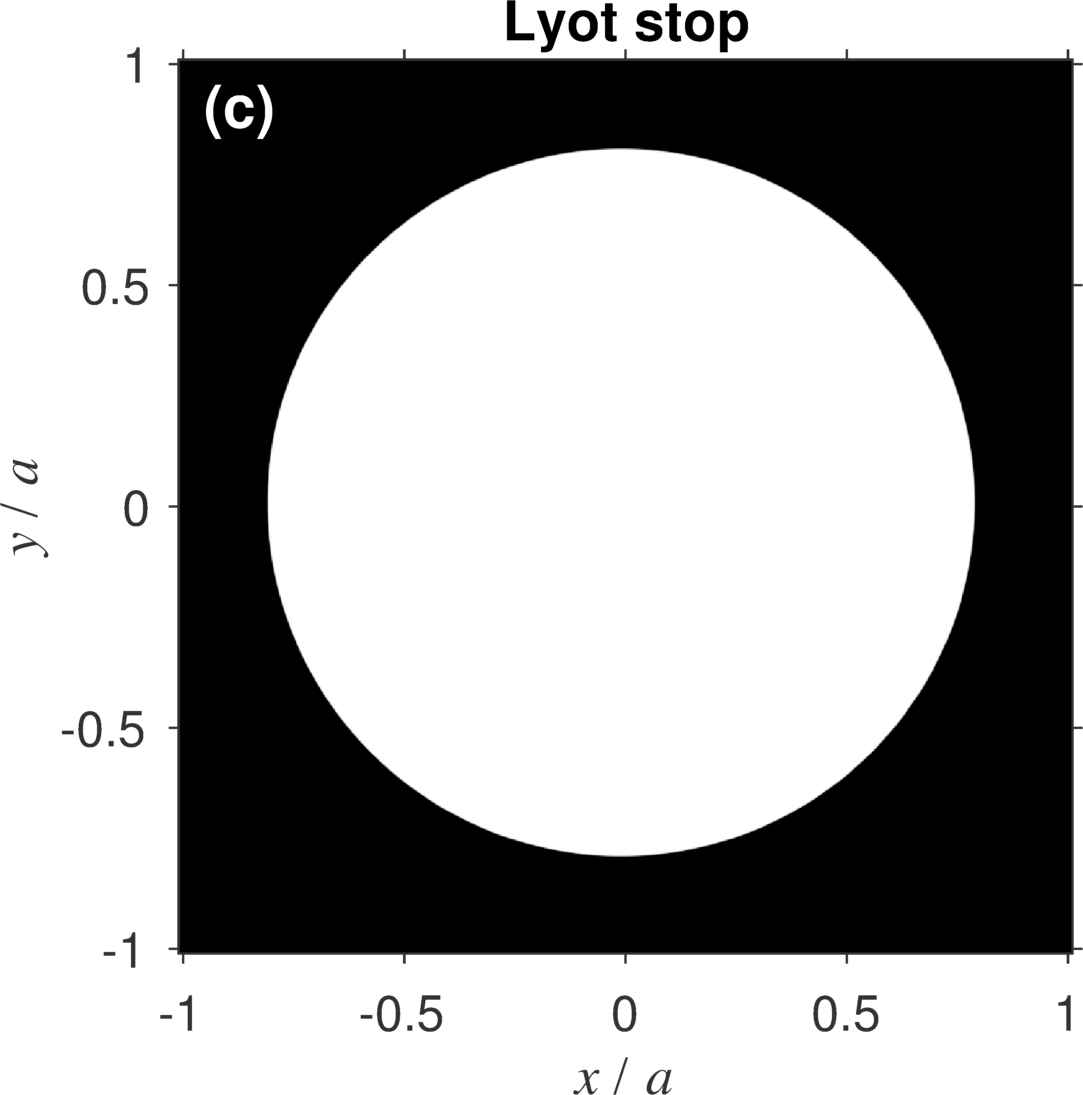}
    \includegraphics[height=5.15cm,trim={0 -0.8cm 0 0},clip]{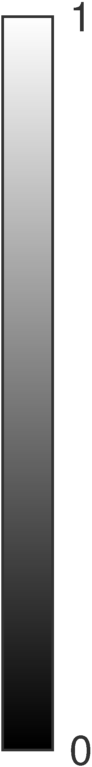}
    \caption{An apodized vortex coronagraph for a 6.5~m HabEx. (a) The image of the primary mirror at the entrance pupil of the coronagraph. (b) The apodizer (squared-magnitude of the desired pupil field). (c) The Lyot stop.}
    \label{fig:apodizedVC_unobs}
\end{figure}

Figure \ref{fig:apodizedVC_unobs}a shows a notional primary mirror with 37 hexagonal segments whose width is $\sim$0.9~m flat-to-flat. The corresponding pupil masks used in the apodized vortex coronagraph are shown in Fig. \ref{fig:apodizedVC_unobs}b,c. The apodizer clips the outer edge of the pupil to make it circular and imparts an amplitude-only apodization pattern on the transmitted or reflected field. Most of the starlight is then diffracted by the vortex outside of the Lyot stop. The small amount of starlight that leaks through the Lyot stop ($\sim$2\%) only contains high-spatial frequencies greater than a specified value $\xi_\text{max}=20$ cycles across the pupil diameter. Thus, in an otherwise perfect optical system, a dark hole appears in the starlight within a 20$\lambda/D$ radius of the star position for all even nonzero values of the vortex charge $l$. We used the Auxiliary Field Optimization (AFO) method (see Jewell et al., these proceedings) to calculate the optimal grayscale pattern\cite{Ruane2016_SPIE}. 

\begin{figure}[t]
    \centering
    \includegraphics[height=0.31\linewidth]{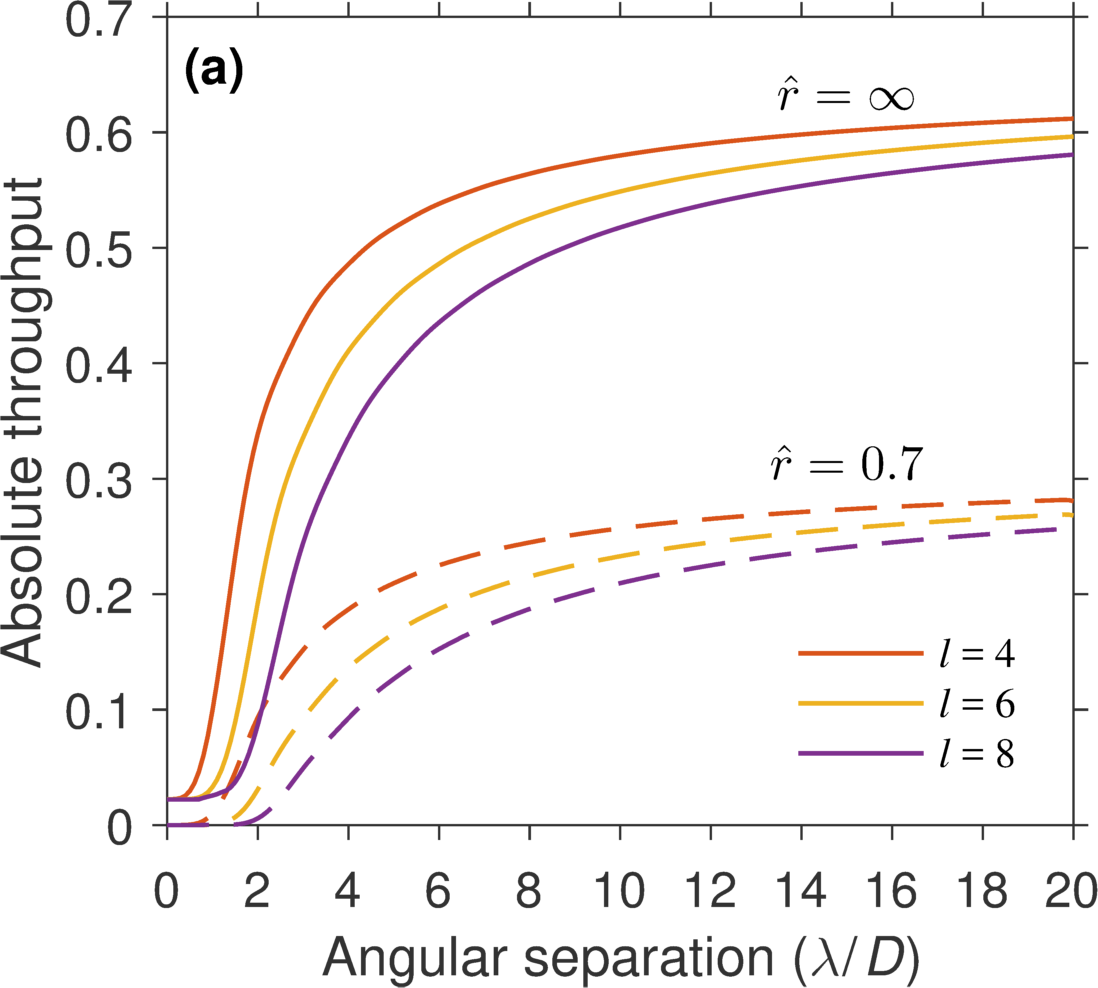}
    \includegraphics[height=0.31\linewidth]{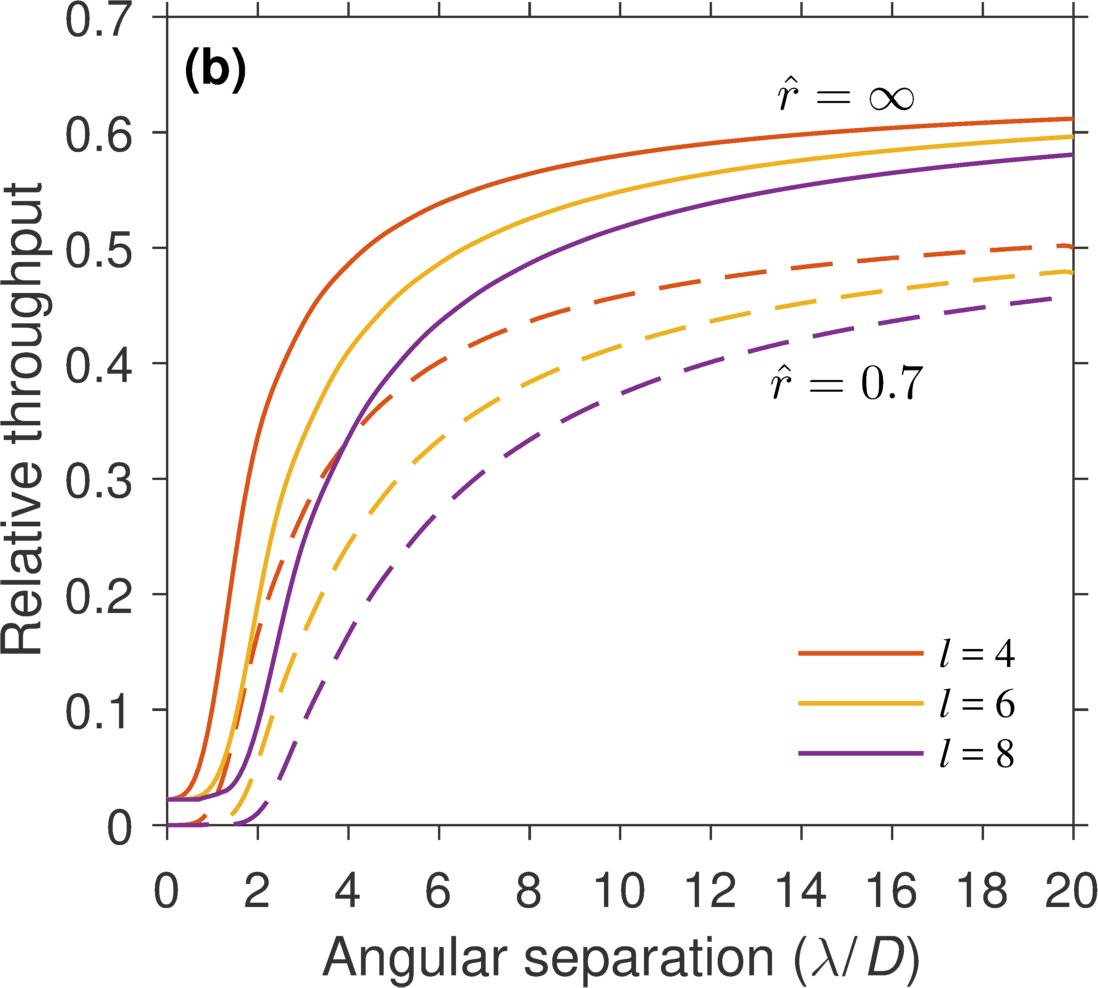}
    \caption{The throughput of the apodized vortex coronagraph with charge 4, 6, and 8 focal plane masks. (a) Absolute throughput. The fraction of total planet light that falls within $\hat{r}\lambda/D$ of the planet position, assuming an otherwise perfect optical system. (b) Relative throughput. The fraction of planet light that falls within $\hat{r}\lambda/D$ of the planet position compared to case with the coronagraph masks removed.}
    \label{fig:Thpt_apodVC_unobs}
\end{figure}

The throughput of the coronagraph with various focal plane vortex masks is shown in Figure \ref{fig:Thpt_apodVC_unobs}. We report both the absolute throughput $\eta_p$ and relative throughput $\eta_p/\eta_\text{tel}$ within a circular region of interest of radius $\hat{r}\lambda/D$ centered on the planet position, where $\eta_\text{tel}$ represents the throughput of the telescope with the coronagraph masks removed. After the coronagraph, $\sim60\%$ of the total energy from an off-axis source remains. Less than 30$\%$ of the total energy appears within $0.7\lambda/D$ of the planet position, including losses from the apodizer and broadening of the point spread function by the undersized pupil mask and Lyot stop. Approximately 50\% of the planet light remains within $0.7\lambda/D$ compared to the point spread function with the coronagraph masks removed. Other than a loss in throughput, the apodized version shares most of the same performance characteristics as the conventional vortex coronagraph. 

\subsection{Sensitivity to low-order aberrations and the angular size of stars}

\begin{figure}
    \centering
    \includegraphics[width=0.8\linewidth,trim={0 -0.11cm 0 0},clip]{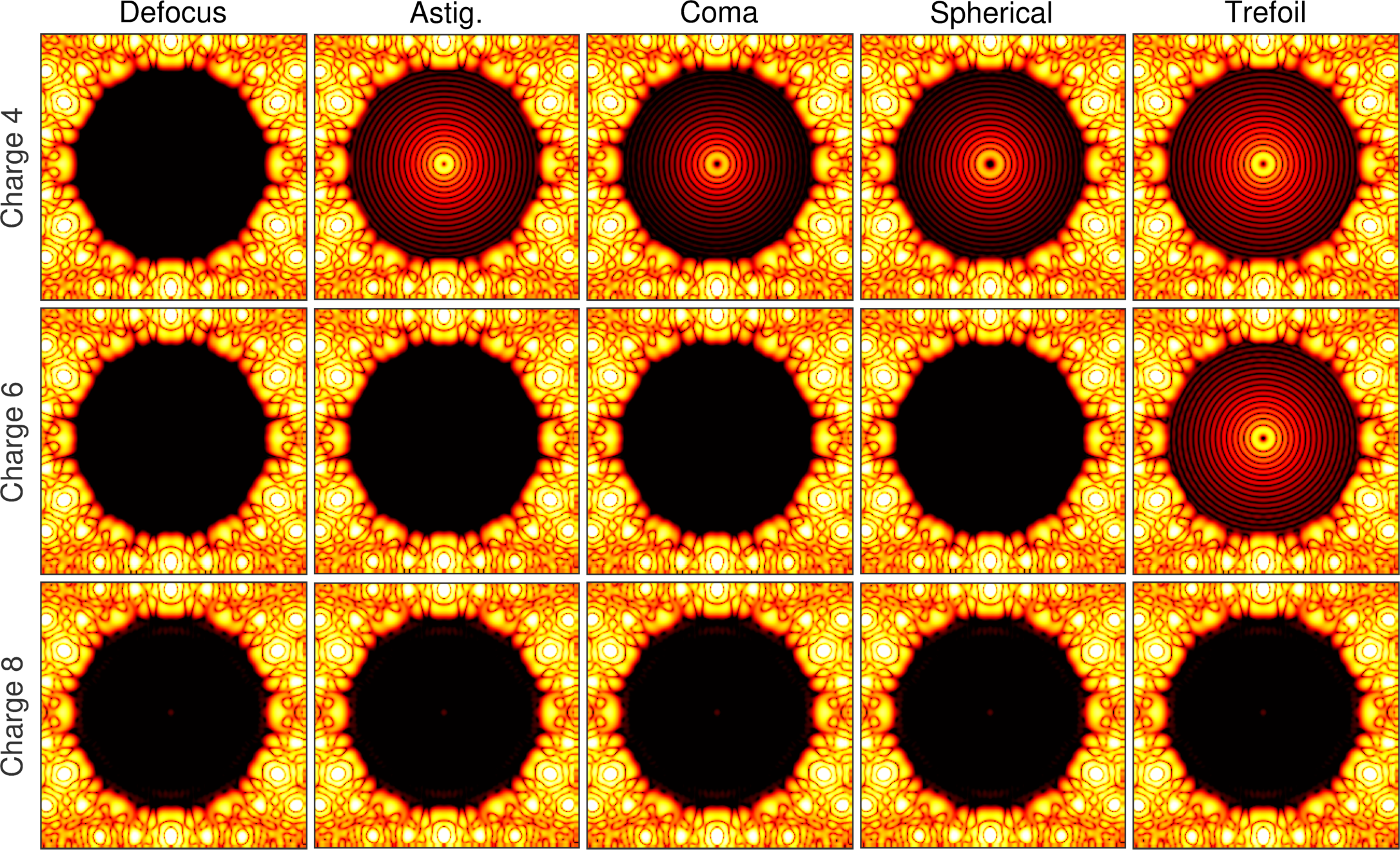}
    \includegraphics[height=8.2cm,trim={0 0 0 0},clip]{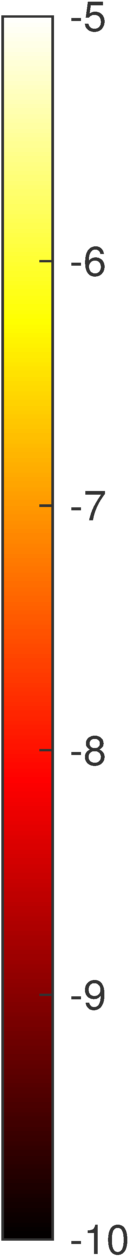}
    \caption{The sensitivity of an apodized vortex coronagraph to low order aberrations on an off-axis, segmented telescope. Log irradiance owing to $\lambda/1000$ rms wavefront error in each mode, normalized to the peak value with the coronagraph masks removed. The dark zone has an angular diameter of 40$\lambda/D$.}
    \label{fig:apodizedVC_unobs_Zsens}
\end{figure}

Assuming the telescope is off-axis and unobstructed, the leaked stellar irradiance in the presence of low-order aberrations appears identical to the monolithic case, up to a radius of $\xi_\text{max} \lambda F^{\#}$ (see Fig. \ref{fig:apodizedVC_unobs_Zsens}). However, to maintain a fixed raw contrast threshold, the wavefront error requirements presented in Table \ref{tab:loworder} scale as $1/\sqrt{\eta_p}$. For example, with a relative throughput of $50\%$, a requirement of 10~pm~rms in Table \ref{tab:loworder} becomes 7~pm~rms. For the sake of brevity, we have not included an updated wavefront error requirement table here. 

\begin{figure}
    \centering
    \includegraphics[width=0.8\linewidth,trim={0 -0.11cm 0 0},clip]{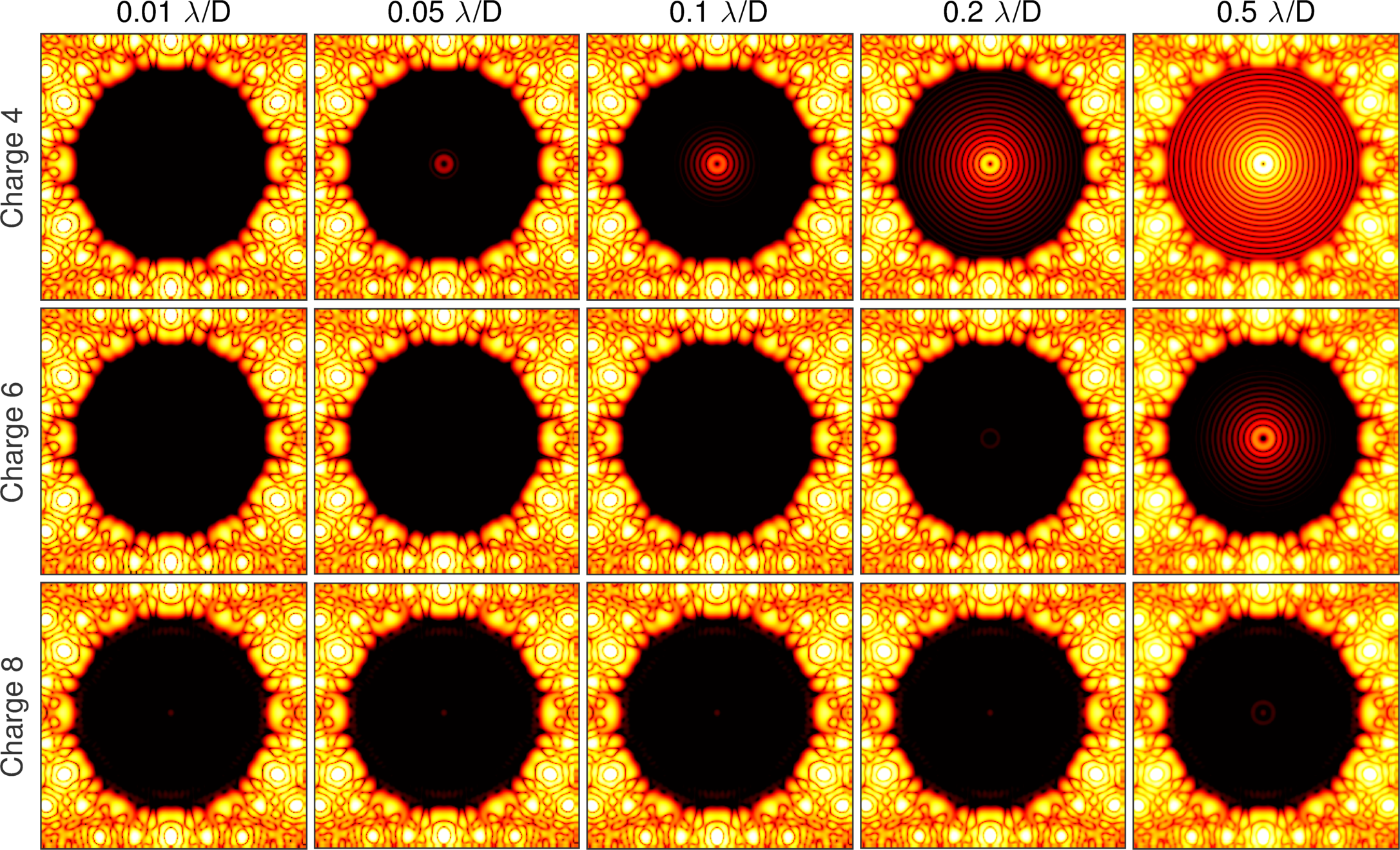}
    \includegraphics[height=8.2cm,trim={0 0 0 0},clip]{colorbarsforfig5.png}
    \caption{The sensitivity of an apodized vortex coronagraph to stellar angular diameter on an off-axis, segmented telescope. Log stellar irradiance, normalized to the peak value with the coronagraph masks removed. The dark zone has a diameter of 40$\lambda/D$. The simulation is monochromatic, but applies to all wavelengths.}
    \label{fig:apodizedVC_unobs_FiniteStarSens}
\end{figure}

The stellar leakage due to the angular size of stars is also equivalent to a vortex coronagraph without an apodizer. Figure \ref{fig:apodizedVC_unobs_FiniteStarSens} shows the leaked starlight as a function of stellar angular size. A charge 4 is sufficient to suppress stars $\lesssim0.01\lambda/D$ in diameter. A charge 6 or 8 may be used to minimize $\eta_s/\eta_p^2$ in the case of a larger star, such as Alpha Centauri A whose angular diameter is 8.5~mas or $\sim$0.5$\lambda/D$ in the visible.

\subsection{Segment co-phasing requirements}

\begin{figure}
    \centering
    \includegraphics[width=0.48\linewidth,trim={0 0 0 0},clip]{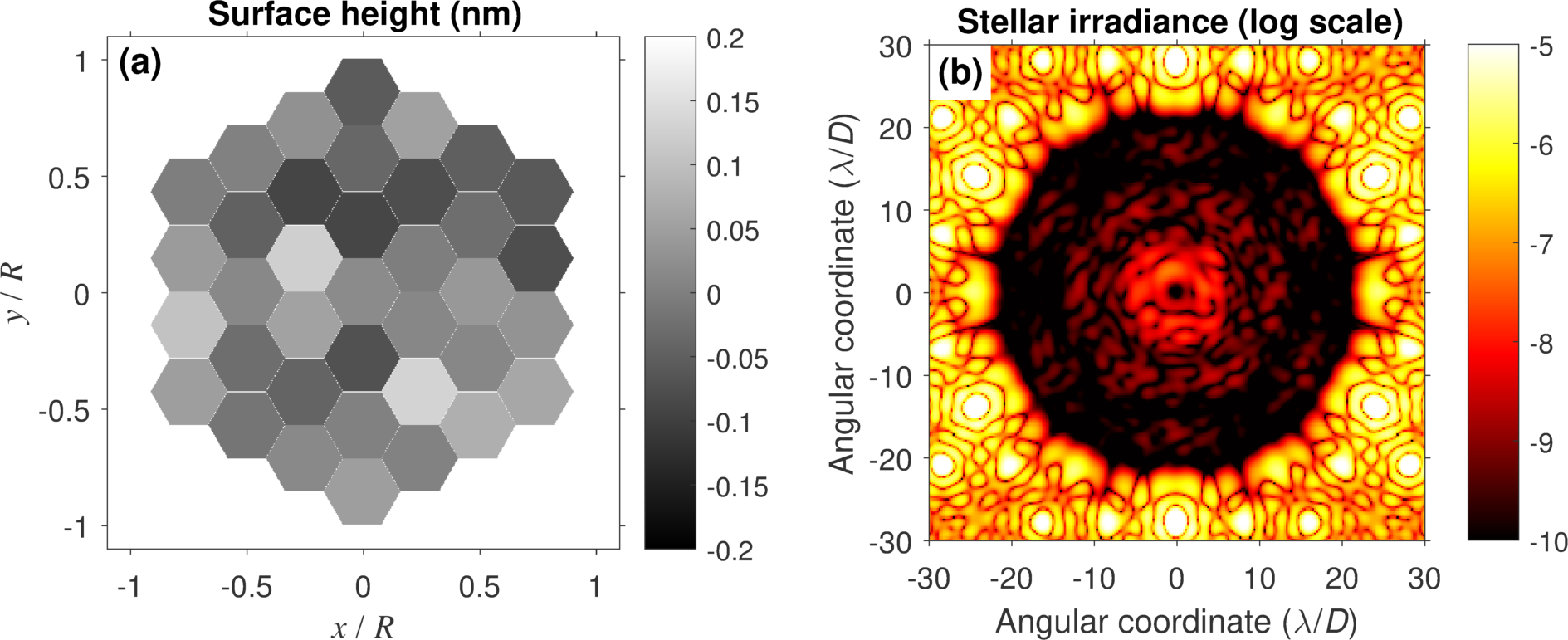}\hfill
    \includegraphics[width=0.48\linewidth,trim={0 0 0 0},clip]{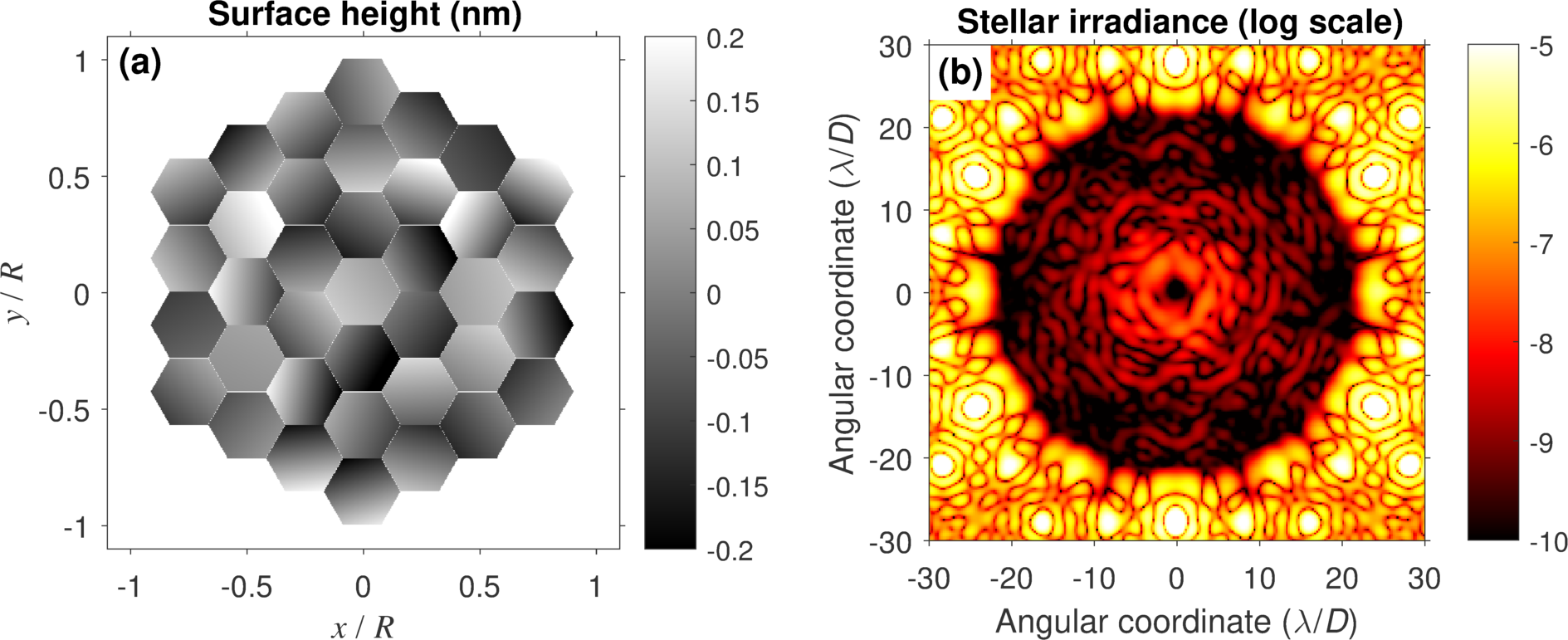}

    \caption{\textit{left:} (a) Example wavefront with 100~pm rms of random segment piston errors and (b) the corresponding stellar irradiance at $\lambda=450~\text{nm}$. \textit{right:} Same as \textit{left}, but with an additional 0.005~$\lambda/D=71~\mu$as~rms of random tip-tilt errors. }
    \label{fig:segErrs}
\end{figure}

\begin{figure}
    \centering
    \includegraphics[height=0.31\linewidth,trim={0 0 0 0},clip]{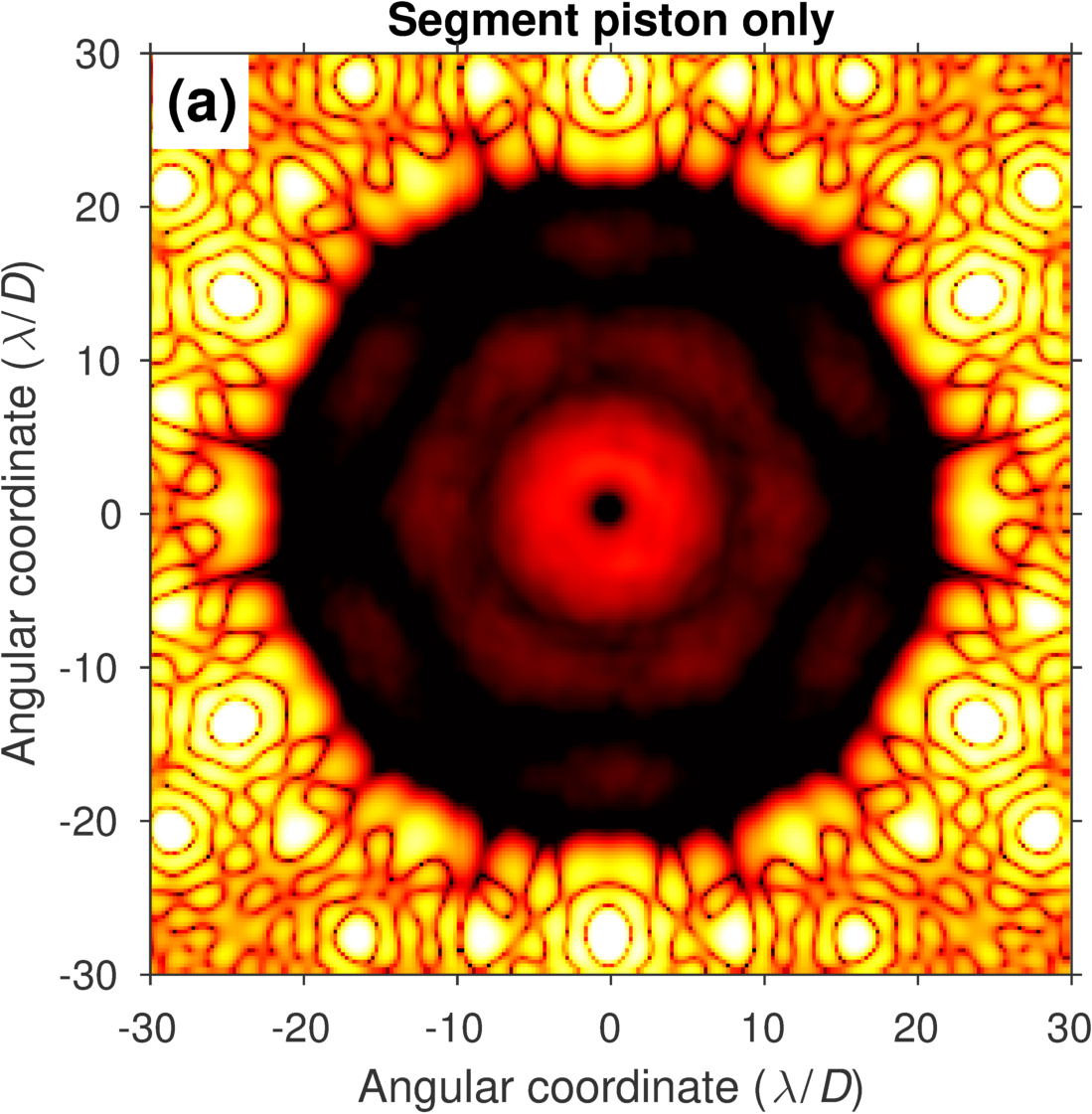}
    \includegraphics[height=0.31\linewidth,trim={1.1cm 0 0 0},clip]{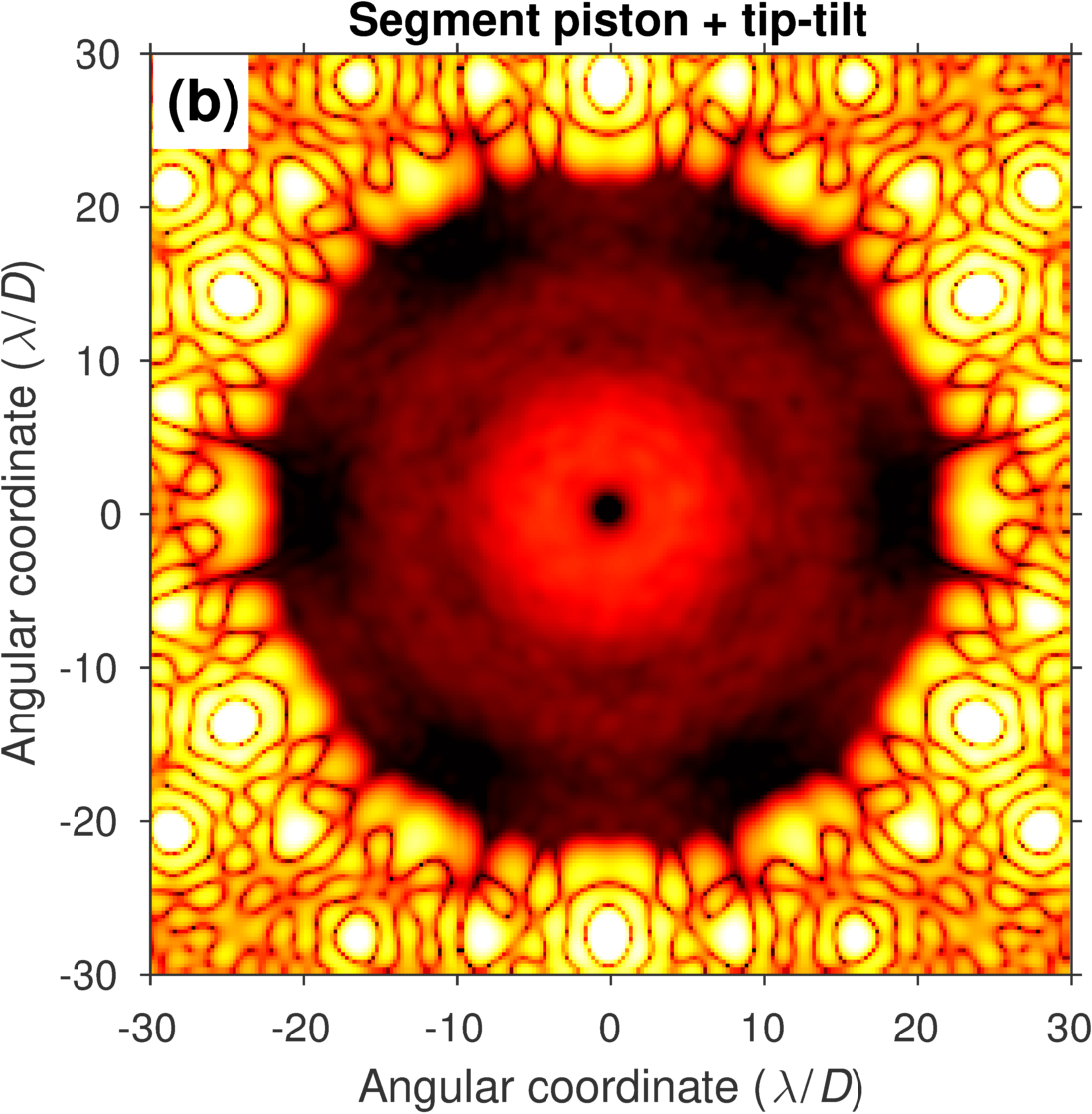}
    \includegraphics[height=5.2cm,trim={0 -0.63cm 0 0},clip]{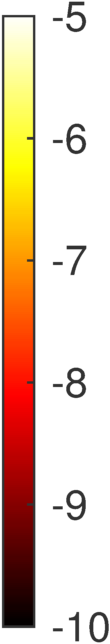}
    \includegraphics[height=0.3\linewidth,trim={0 0 0 0},clip]{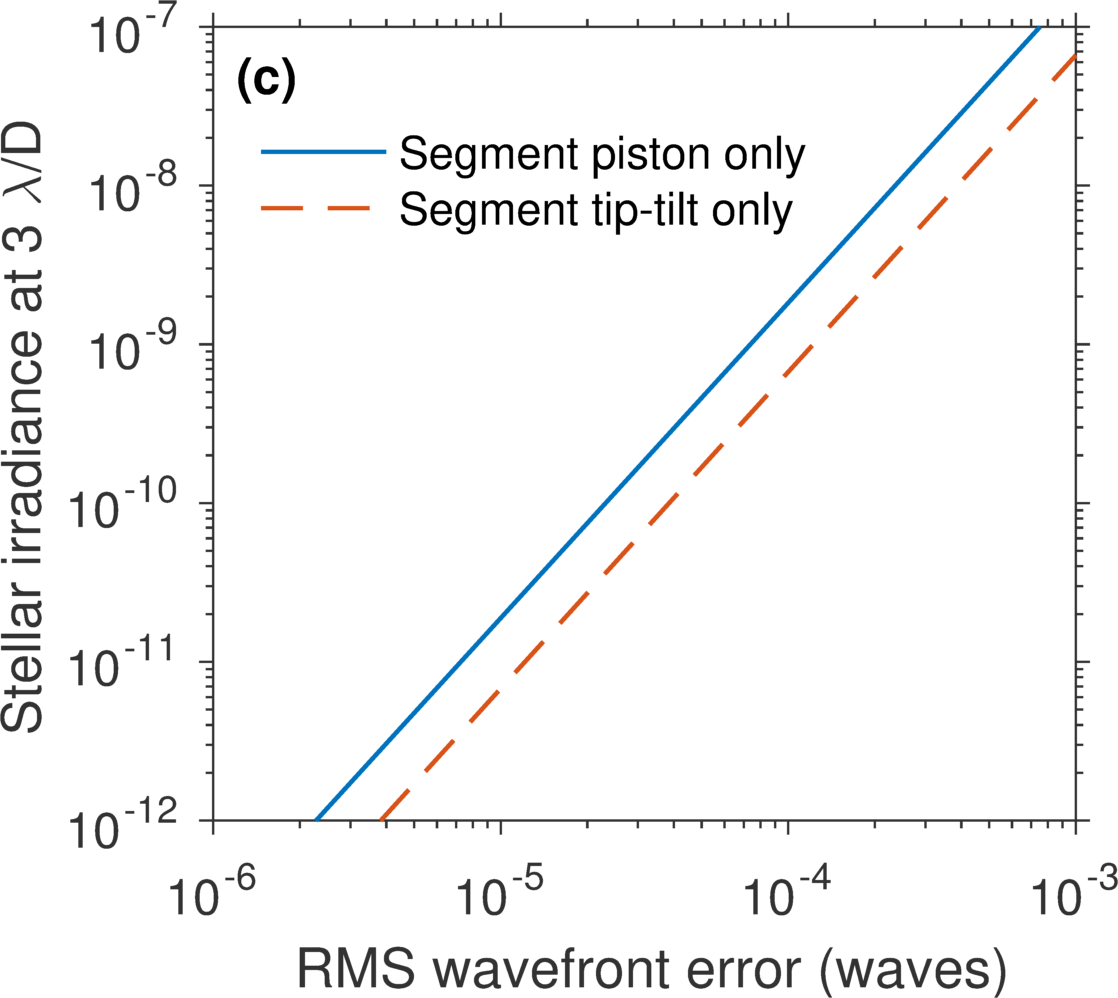}
    \caption{(a)-(b) Time average over many realizations of leaked stellar irradiance at $\lambda=450$~nm due to (a) 100~pm rms of random segment piston and (b) with an additional 0.005~$\lambda/D=71~\mu$as~rms tip-tilt error. (c) Dependence of stellar irradiance at 3~$\lambda/D=43$~mas on the rms wavefront error segment piston and tip-tilt. The simulation is monochromatic, but applies to all wavelengths.}
    \label{fig:segErrs2}
\end{figure}

A major challenge for exoplanet imaging with a segmented telescope will be to keep the mirrors co-aligned throughout observations. As shown in Fig.~\ref{fig:segErrs}, small segment motions in piston and tip-tilt cause speckles to appear in the dark hole which may be difficult to calibrate and will likely contribute to both $\eta_s$ and $n_2$. Figure~\ref{fig:segErrs2}a-b shows the time average over many realization of the errors shown in Fig. \ref{fig:segErrs} drawn from Gaussian distributions for both piston and tip-tilt with standard deviations of 100 pm and 0.005~$\lambda/D=71~\mu$as~rms. When the mirror segments have random piston errors only, the resulting distribution of light resembles the diffraction pattern of a single segment (Fig.~\ref{fig:segErrs2}a). Random segment tip-tilt error tend to spread the leaked starlight to larger separations (Fig.~\ref{fig:segErrs2}b). However, the leaked starlight is well approximated by a similar second order power law in both cases (Fig.~\ref{fig:segErrs2}c), which yields a wavefront error requirement of $\sim$10~pm~rms, similar in magnitude to unsuppressed low-order modes. On the other hand, if the primary mirror segments undergo a coordinated movement that resembles a low-order Zernike polynomial $Z_n^m$, the amount of leaked starlight would be significantly smaller if $l>n+|m|$ and the tolerance to such a motion would be considerably relaxed. 

\section{CENTRALLY-OBSCURED, SEGMENTED TELESCOPE} \label{sec:onaxis_seg} 

The LUVOIR mission concept is a 9-15~m centrally-obscured, segmented telescope which will spectrally characterize tens of earth-like exoplanets using its internal coronagraph. In this section, we present an apodized vortex coronagraph design for LUVOIR and compare its performance with the HabEx vortex coronagraphs. 

\subsection{Apodizers}

Apodized vortex coronagraphs for centrally obscured telescopes differ significantly from those that are optimized for off-axis telescopes. The apodizer must include a sharp radial feature that generates additional diffraction which destructively interferes with diffracted light the secondary mirror\cite{Mawet2013_ringapod,Fogarty2017}. Thus, the Lyot stop inner radius must match the radius of the ring feature and only transmit in the region where the intended destructive interference occurs. The apodizer must be numerically optimized for a single choice of Lyot stop inner radius and vortex mask charge in order to satisfy these additional constraints. 

\begin{figure}
    \centering
    \includegraphics[height=0.31\linewidth]{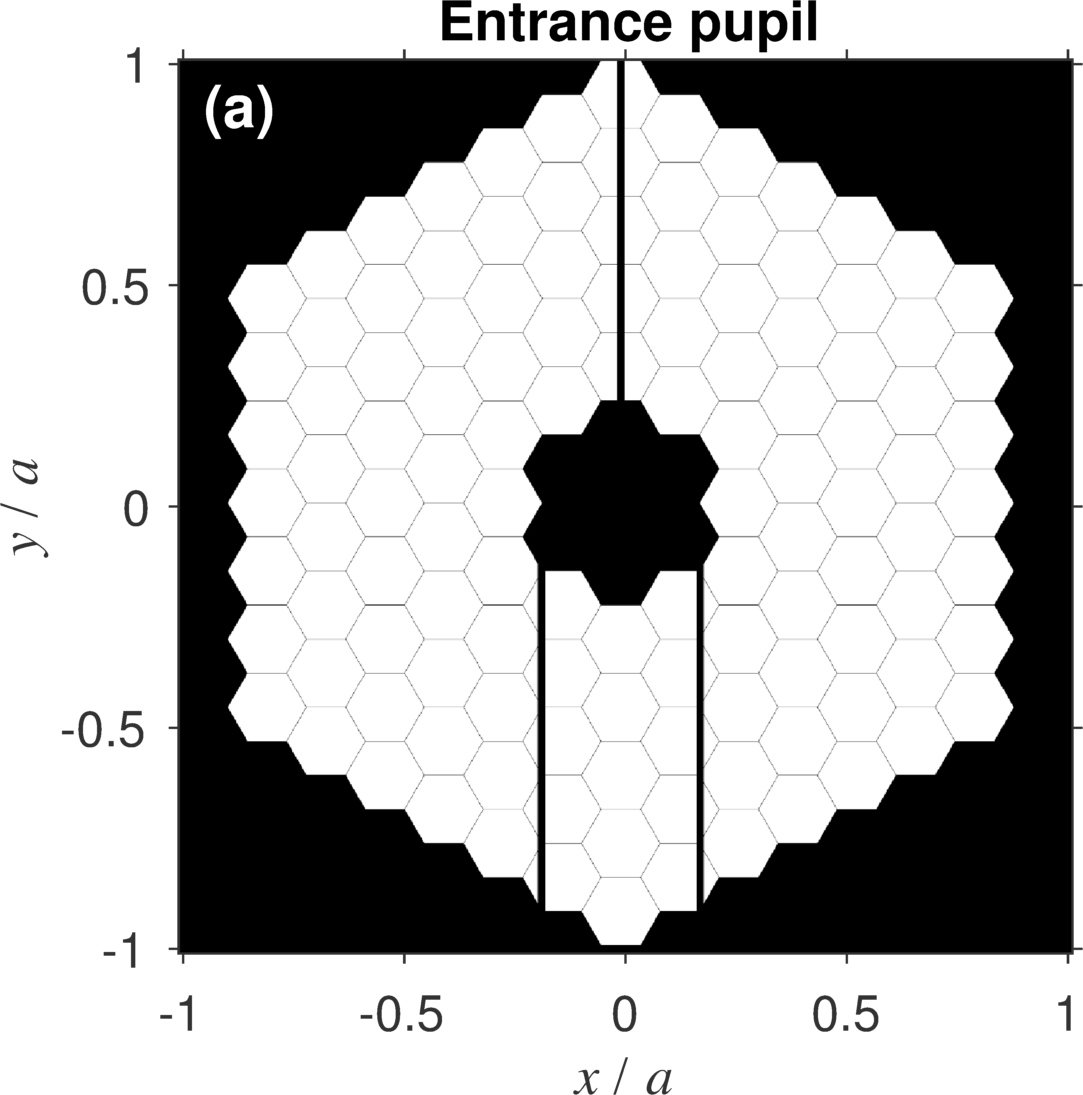}
    \includegraphics[height=0.31\linewidth,trim={1.4cm 0 0 0},clip]{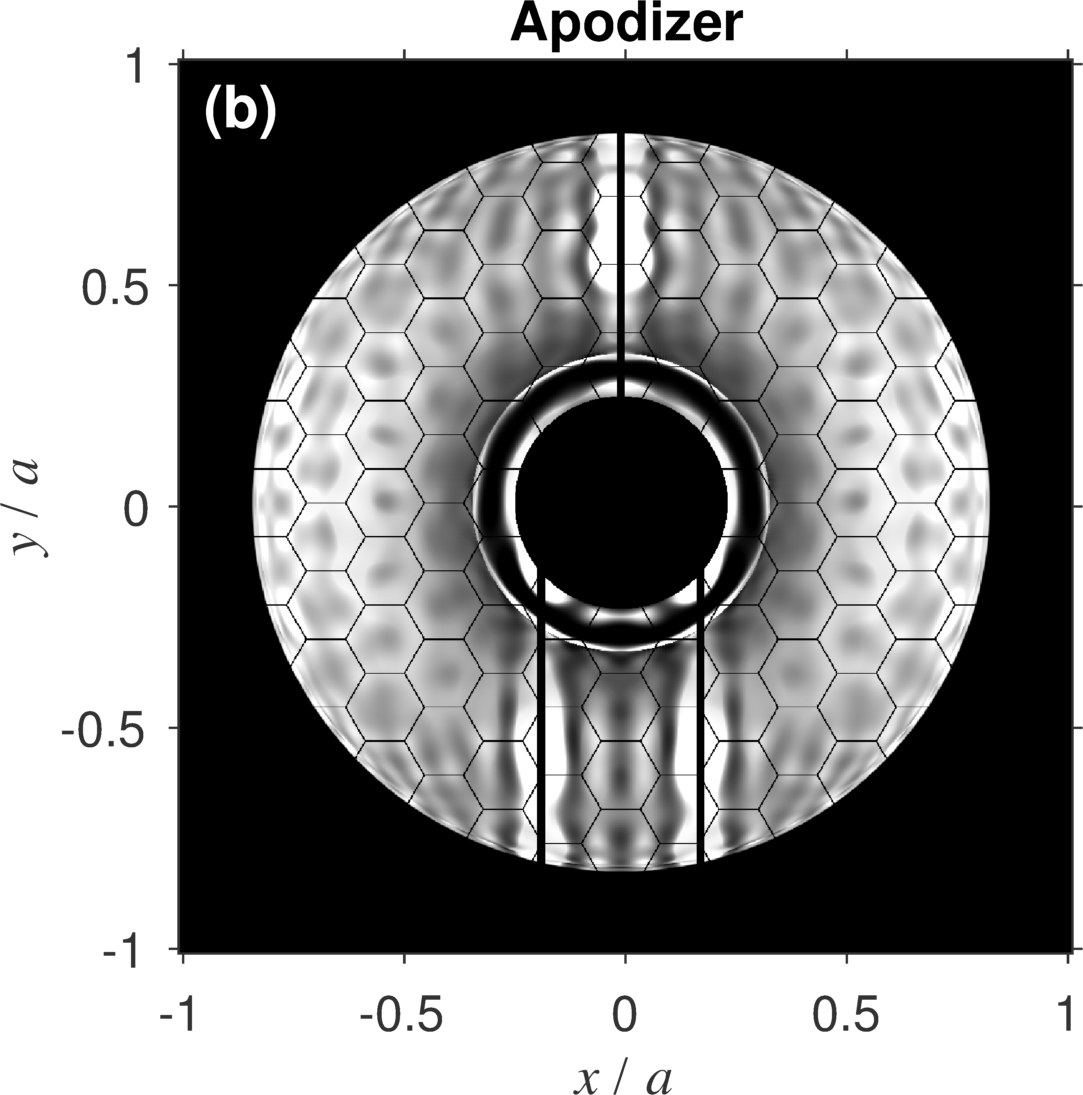}
    \includegraphics[height=0.31\linewidth,trim={1.4cm 0 0 0},clip]{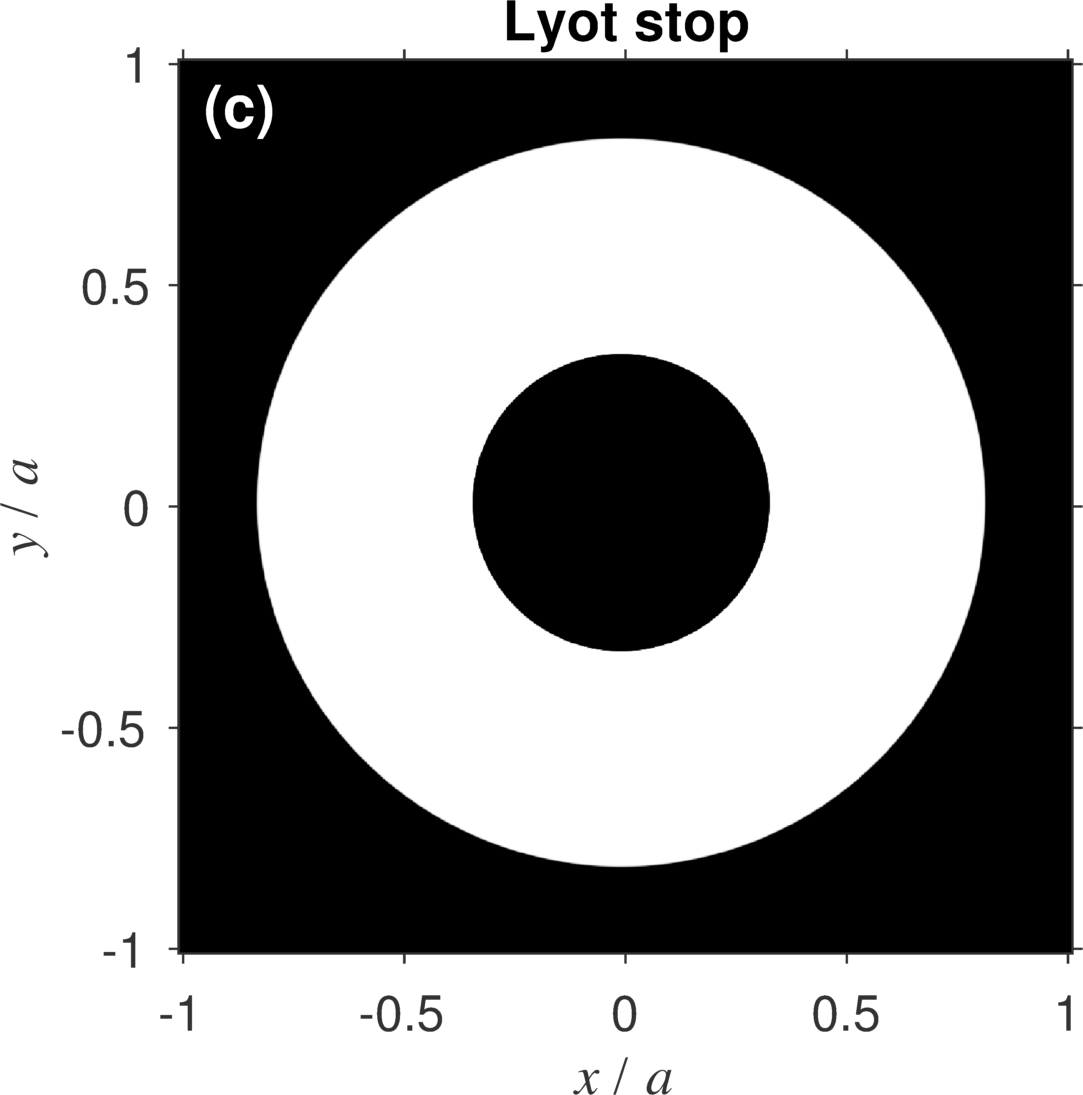}
    \includegraphics[height=5.15cm,trim={0 -0.8cm 0 0},clip]{colorbarsforfig3.png}
    \caption{An apodized vortex coronagraph for a conceptual LUVIOR telescope. (a) The image of the primary mirror at the entrance pupil of the coronagraph. The (b) apodizer (squared-magnitude of the desired pupil field) and (c) Lyot stop for a charge 8 vortex coronagraph.}
    \label{fig:apodizedVC_LUVOIR}
\end{figure}

Figure \ref{fig:apodizedVC_LUVOIR} shows a conceptual design for a LUVOIR vortex coronagraph. The image of the telescope aperture consists of five rings of hexagonal mirrors, a secondary mirror, and vertical spider supports (Fig. \ref{fig:apodizedVC_LUVOIR}a). The apodizer presented here was optimized for a charge 8 vortex focal plane mask (see Fig. \ref{fig:apodizedVC_LUVOIR}a-b). 

\begin{figure}[t]
    \centering
    \includegraphics[height=0.31\linewidth]{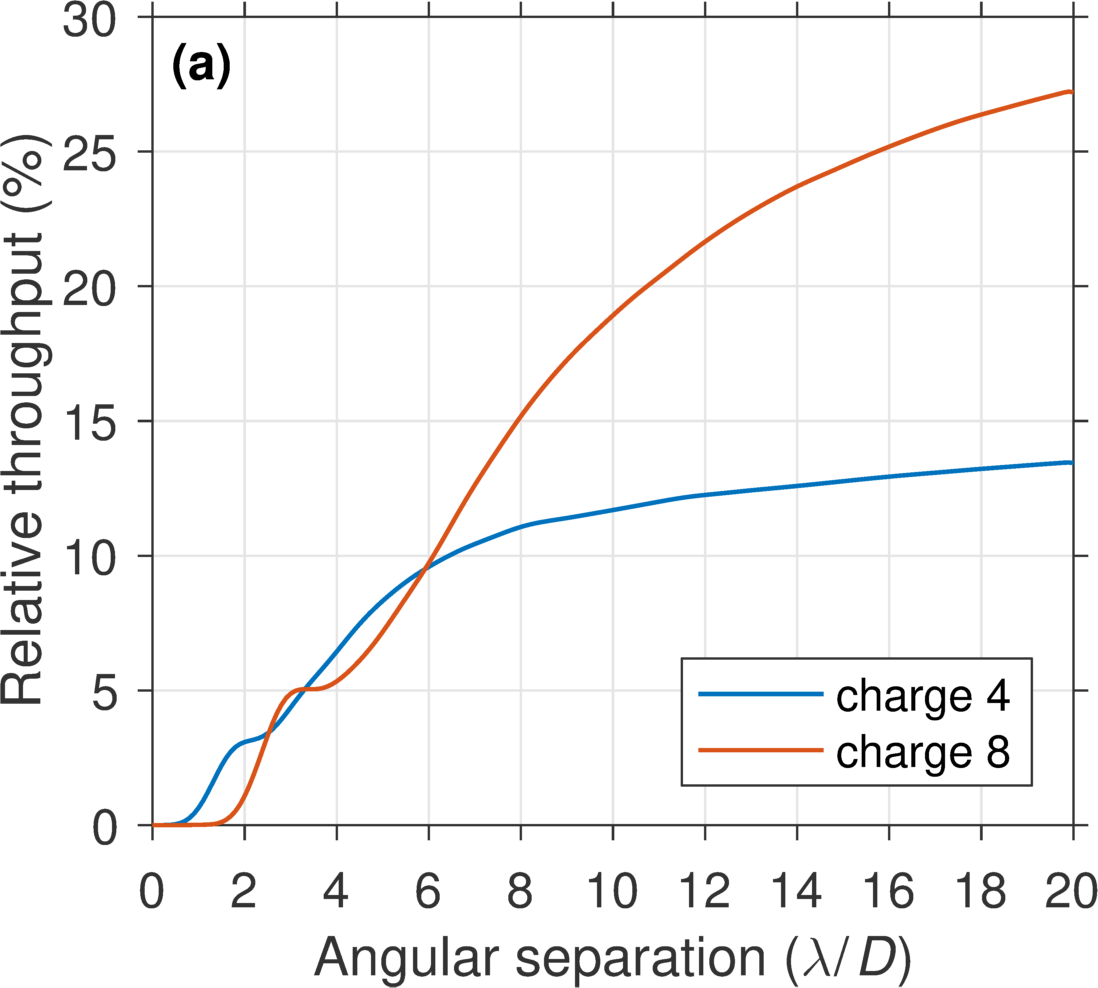}
    \includegraphics[height=0.31\linewidth]{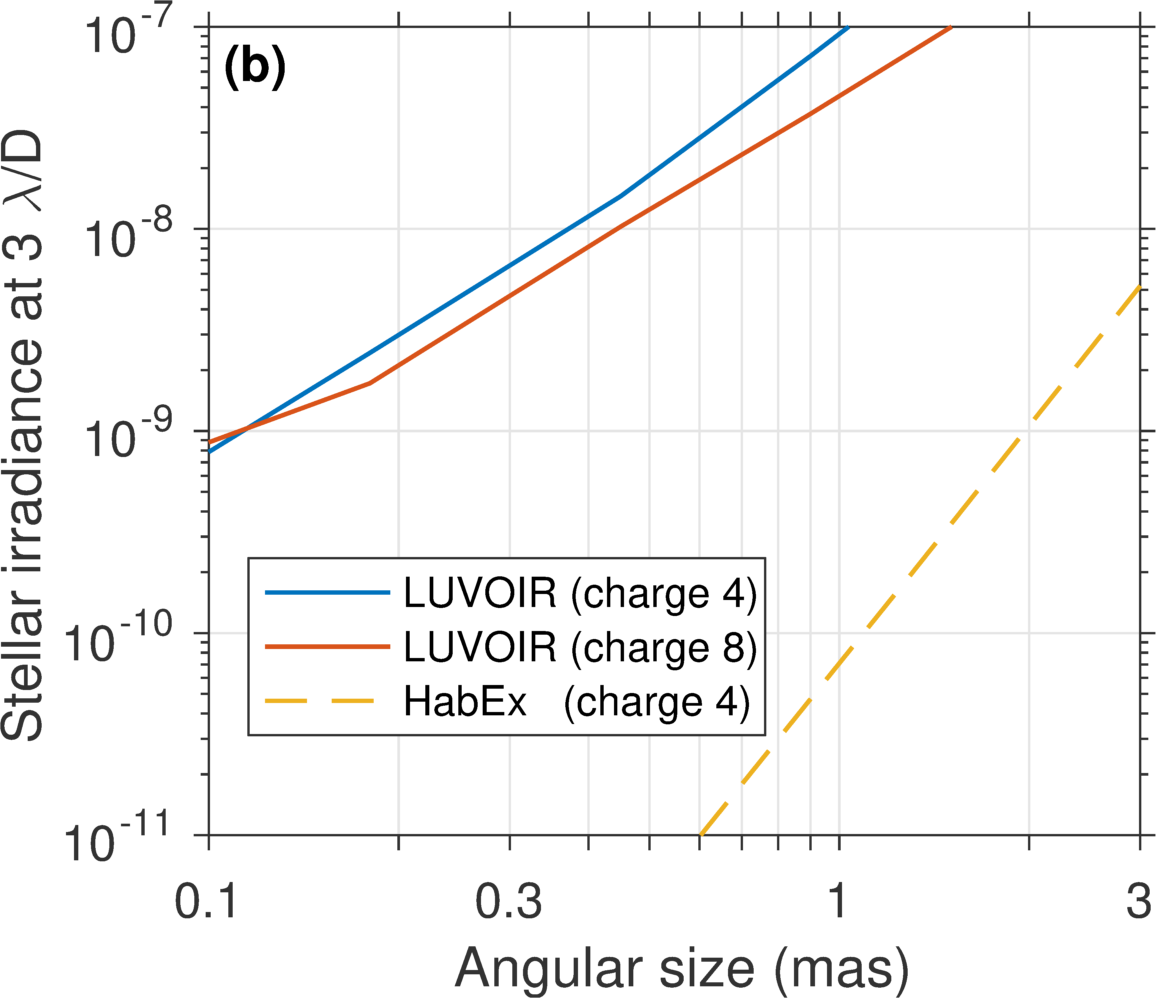}
    \caption{(a) Relative throughput of an apodized vortex coronagraph with charge 4 and 8 focal plane masks within $0.7\lambda/D$ of the planet position, assuming an otherwise perfect optical system. (b) The corresponding sensitivity to the finite size of the star compared to a charge 4 HabEx (with a 4~m off-axis monolith) vortex coronagraph.}
    \label{fig:Thpt_apodVC_LUVOIR}
\end{figure}

Compared to an off-axis telescope, a centrally obscured telescope generally leads to a lower transmission apodizer and lower coronagraph throughput. What is more, the relationship between throughput and $l$ no longer follows a regular trend\cite{Fogarty2017}. For example, Fig. \ref{fig:Thpt_apodVC_LUVOIR}a shows the throughput of a charge 4 and 8 designs. Although the lower charge still provides better throughput at the smallest angular separations, the throughput at larger angular separations is higher for charge 8. The low order aberration sensitivity can also follow similarly anomalous trends. However, comparable low order aberration insensitivity to that of conventional vortex coronagraphs may be achieved by tuning the inner radius of the Lyot stop. 

\subsection{Sensitivity to the size of the star}

The major challenge for designing small inner working angle coronagraphs on centrally obscured telescopes is suppressing leakage owing to the angular size of the star. To illustrate this, Fig. \ref{fig:Thpt_apodVC_LUVOIR}b compares the leaked starlight as a function of stellar size for LUVOIR vortex coronagraphs and a charge 4 vortex coronagraph on a 4~m HabEx telescope. The central obscuration causes the LUVOIR coronagraph to leak significantly more light than the smaller off-axis architecture. The additional leaked starlight increases $\eta_s$ considerably for all positions of interest. 

We compare the resulting performance by calculating the relative exposure time to detect a planet at an angular separation of 50~mas from a 1~mas star. In the visible regime, this separation corresponds to $\sim3~\lambda/D$ with the 4~m HabEx and $\sim7.5~\lambda/D$ with the 15~m LUVOIR. Whereas the relative throughput for a charge 4 HabEx coronagraph is $\sim41\%$, the charge 8 LUVOIR coronagraph is $\sim14\%$. As outlined in section \ref{sec:metrics}, the relative exposure time is
\begin{equation}
    \frac{\Delta t_{4\text{m}}}{\Delta t_{15\text{m}}}=\frac{\eta_{s,4\text{m}}}{\eta_{s,15\text{m}}}\left(\frac{\eta_{p,15\text{m}}}{\eta_{p,4\text{m}}}\right)^2\frac{A_{15\text{m}}}{A_{4\text{m}}},
\end{equation}
where the subscripts 4m and 15m indicate the aperture size each quantity represents. With the throughput loss introduced by the coronagraph masks, ensuring superior performance of the LUVOIR coronagraph requires $\eta_{s,15\text{m}} \lesssim 1.2~\eta_{s,4\text{m}}$. Thus, the current apodized vortex coronagraph designs are far from enabling the full scientific potential of the LUVOIR missions. However, other designs have been proposed that perform well on the LUVOIR architecture for typical star sizes, such as apodized Lyot coronagraphs with binary amplitude pupil masks known as ``shaped pupils."\cite{Zimmerman2016} 

Despite the relatively poor raw suppression provided by the apodized vortex coronagraph on LUVOIR, such a design may be incorporated in the instrument to enable spectral characterization of planets in the infrared channels. The modest amount of throughput at small angular separations may be necessary if the planet was detected near the inner working angle in the visible (currently $\sim 4~\lambda/D$).


\subsection{Segment co-phasing requirements}

The segment co-phasing errors requirements are roughly equivalent to the off-axis case. Again, if a fixed "raw contrast" is desired, the wavefront error requirements scale with $1/\sqrt{\eta_p}$. The smaller mirror segments with respect to the pupil diameter provides a slight relaxation of requirements owing to broader diffraction pattern in the image plane owing to a single segment. However, aside from these minor corrections, the stability requirement in terms of RMS wavefront error is approximately 10~pm~rms.

\section{CONCLUSION AND FUTURE OUTLOOK} \label{sec:conc} 

Apodized vortex coronagraphs are a very promising solution for imaging earth-like exoplanets with the HabEx and LUVOIR decadal mission concepts. The off-axis design of the HabEx telescope allows for the best performance in terms of throughput, inner working angle, and robustness to aberrations. In the case of on-axis, obscured telescopes, the apodized vortex coronagraph designs presented here become far more sensitive to the finite size of stars, which introduces additional photon noise and significantly increases exposure times for detection. 

Further work is required to optimize the LUVOIR coronagraphs designs to suppress sources with angular diameters of $\sim$1 mas. The sensitivity losses owing to leakage from the finite size of star may be compensated by drastically improving throughput. One potential pathway for doing so is to use beam shaping techniques \cite{Fogarty2017}. In addition, high dispersion coronagraphy methods may be applied to reduce the exposure time needed to characterize the atmospheres of earth-like planets \cite{Wang2017,Mawet2017b}. Experimental demonstrations of apodized vortex coronagraphs and high dispersion coronagraphy techniques are underway on the High Contrast Spectroscopy Testbed for Segmented Telescopes (HC(ST)$^2$) at Caltech's Exoplanet Technology (ET) Laboratory (see Delorme et al., these proceedings).

\appendix

\section{Zernike polynomials}

The Zernike polynomials \cite{Zernike1934} may be written as 
\begin{equation}
    Z_n^m\left(r/a,\theta\right) = R_n^{|m|}\left(r/a\right)
    \begin{cases} 
        \cos\left(m\theta\right) & m \geq 0 \\
        \sin\left(|m|\theta\right) & m<0
    \end{cases}
    ,\;\;\;\;\;r\le a,
\label{eq:Zpupil_appendix}
\end{equation}
where $R_n^m(r/a)$ is the radial Zernike polynomial given by 
\begin{equation}
R_n^m(r/a) = \sum_{k=0}^{\frac{n-m}{2}}\frac{(-1)^k(n-k)!}{k!\left(\frac{n+m}{2}-k\right)!\left(\frac{n-m}{2}-k\right)!}(r/a)^{n-2k},\;\;\;\;\;r/a\le 1,
\end{equation}
where $n-m$ is even. The indices $n$ and $m$ are integers respectively known as the degree and azimuthal order. The first few radial polynomials are: $R_0^0=1$, $R_1^1=r/a$, $R_2^0=2(r/a)^2-1$, $R_2^2=(r/a)^2$, $R_3^1=3(r/a)^3-2(r/a)$, $R_3^3=(r/a)^3$. 

\section{Mechanism for low-order aberration insensitivity}

An isolated phase aberration is written
\begin{equation}
P(r,\theta) = \exp \left[i c_{nm} Z_n^m(r/a,\theta)\right],\;\;\;\;\;r\le a,
\end{equation}
where $i=\sqrt{-1}$ and $c_{n,m}$ is the Zernike coefficient. Assuming small wavefront errors (i.e. $c_{nm}\ll$ 1 rad), the field in the pupil may be approximated to first order via its Taylor series expansion:
\begin{equation}
P(r,\theta) \approx 1 + i c_{nm} Z_n^m(r/a,\theta),\;\;\;\;\;r\le a.
\label{eq:Ztaylor}
\end{equation}
For convenience, we choose to use the set real-valued of Zernike polynomials described by
\begin{equation}
    Z_n^m(r/a,\theta) = R_n^{|m|}(r/a)q_m(\theta)
    ,\;\;\;\;\;r\le a,
\label{eq:Zpupil}
\end{equation}
where $R_n^m\left(r/a\right)$ are the radial polynomials described in Appendix A and 
\begin{equation}
    q_m(\theta) = 
    \begin{cases} 
        \cos(m\theta) & m \geq 0 \\
        \sin(|m|\theta) & m<0
    \end{cases}
    .
\label{eq:q}
\end{equation}
The field transmitted through the vortex mask, owing to an on-axis point source, is given by the product of $\exp\left(il\phi\right)$ and the optical Fourier transform (FT) of Eq. \ref{eq:Ztaylor}:
\begin{equation}
F_{nml}(\rho,\phi)=\left[f_{00}(\rho,\phi) + ic_{nm}f_{nm}(\rho,\phi) \right]e^{il\phi},
\label{eq:ZPSF}
\end{equation}
where
\begin{equation}
f_{nm}(\rho,\phi) =\frac{k a^2}{f}\frac{J_{n+1}\left( k a \rho/f\right)}{k a \rho/f} q_m(\phi),
\label{eq:ZPSF2}
\end{equation}
$\rho$ is the radial polar coordinate in the focal plane, $k=2\pi/\lambda$, and $f$ is the focal length. The field in the subsequent pupil plane (i.e. just before the Lyot stop), $E_{lnm}$, is given by the FT of Eq. \ref{eq:ZPSF}. The first term, $f_{00}(\rho,\phi)$, is the common Airy pattern, which diffracts completely outside of the Lyot stop for all even nonzero values of $l$. In this case, the field becomes\cite{Carlotti2009}
\begin{equation}
E_{l,\mathrm{Airy}}(r,\theta)=
    \begin{cases} 
        0 & r \le a \\
        \frac{a}{r}R_{|l|-1}^1(\frac{a}{r})e^{il\theta} & r>a
    \end{cases}.
\label{eqn:El00}
\end{equation}
More generally, the full pupil field is given by
\begin{equation}
E_{nml}(r,\theta)=E_{l,\mathrm{Airy}}(r,\theta) + ic_{nm}g_{nml}(r,\theta),
\label{eqn:Elnm}
\end{equation}
where
\begin{equation}
g_{nml}(r,\theta)=\frac{k a}{2 f}e^{il\theta}
    \begin{cases} 
        (-1)^m e^{im\theta}\mathcal{W}_{n+1}^{l+m}(r)+e^{-im\theta}\mathcal{W}_{n+1}^{l-m}(r) & m \ge 0 \\
        i\left[(-1)^{m+1} e^{im\theta}\mathcal{W}_{n+1}^{l+m}(r)+e^{-im\theta}\mathcal{W}_{n+1}^{l-m}(r)\right] & m<0
    \end{cases},
\label{eqn:glnm}
\end{equation}
and $\mathcal{W}_p^q(r)$ is a special case of the Weber-Schafheitlin integral (see Appendix C):
\begin{equation}
\begin{split}
\mathcal{W}_p^q(r)=&W_{p,q,0}(r;{k a \rho}/{f},{k r \rho}/{f})\\
=&\int\limits_{0}^{\infty }{J_p\left({k a \rho}/{f}\right)J_q\left( {k r \rho}/{f}\right)d\rho}.
\end{split}
\end{equation}

\section{Weber-Schafheitlin integrals}

The pupil functions generated by vortex coronagraphs are a subset of solutions of the discontinuous integral of Weber and Schafheitlin \cite{Watson1922}, which in its conventional form is written
\begin{equation}
W_{\nu,\mu,\lambda}(t;\alpha,\beta)=\int_0^\infty{ \frac{J_{\nu}(\alpha t)J_{\mu}(\beta t)}{t^\lambda} dt},
\label{eqn:WSintegral}
\end{equation}
where $\nu,\mu,\lambda$ are integers and $\alpha$ and $\beta$ are constants. The integral is convergent provided $\nu+\mu-\lambda \ge 0$ and $\lambda \ge 0$. If $0<\alpha<\beta$,
\begin{align}
W_{\nu,\mu,\lambda}(t;\alpha,\beta)=&\frac{\alpha^{\nu}\Gamma \left( \frac{\nu + \mu - \lambda + 1}{2} \right)}{2^{\lambda}\beta^{\nu-\lambda+1}\Gamma \left( \frac{-\nu + \mu + \lambda + 1}{2} \right) \Gamma(\nu+1)} \\ 
&\times \prescript{}{2}F_{1}\left( \frac{\nu + \mu - \lambda + 1}{2},\frac{\nu - \mu - \lambda + 1}{2};\nu+1;\frac{\alpha^2}{\beta^2} \right),
\end{align}
where $\Gamma(~)$ is the gamma function and $\prescript{}{2}F_{1}(~)$ is a hypergeometric function \cite{Gradshteyn}. On the other hand, if $0<\beta<\alpha$
\begin{align}
W_{\nu,\mu,\lambda}(t;\alpha,\beta)=&\frac{\beta^{\nu}\Gamma \left( \frac{\nu + \mu - \lambda + 1}{2} \right)}{2^{\lambda}\alpha^{\nu-\lambda+1}\Gamma \left( \frac{\nu - \mu + \lambda + 1}{2} \right) \Gamma(\nu+1)} \\ &\times \prescript{}{2}F_{1}\left( \frac{\nu + \mu - \lambda + 1}{2},\frac{-\nu + \mu - \lambda + 1}{2};\mu+1;\frac{\beta^2}{\alpha^2} \right).
\end{align}
Integrals with the form of Eqn. \ref{eqn:WSintegral}, namely a product of Bessel functions, appear in the output function integral in cases where the input function is circular or may be described by a Zernike polynomial in amplitude \cite{Ruane2015_SPIE}.

\acknowledgments     
We thank Matthew Bolcar (NASA GSFC) for providing information regarding the LUVOIR telescope concept as well as Laurent Pueyo (STScI) and Kevin Fogarty (JHU) for fruitful discussion regarding optimal vortex coronagraph designs. G. Ruane is supported by an NSF Astronomy and Astrophysics Postdoctoral Fellowship under award AST-1602444. This work was supported by the Exoplanet Exploration Program (ExEP), Jet Propulsion Laboratory, California Institute of Technology, under contract to NASA.


\bibliography{RuaneLibrary}   
\bibliographystyle{spiebib}   

\end{document}